\documentclass[preprint]{aastex}
\usepackage{graphicx}
\usepackage{url}
\usepackage{color}
\usepackage{natbib}

\shorttitle{The PRL Precision RV Spectrograph}
\shortauthors{Chakraborty et al.}
\slugcomment {accepted for publication in PASP}

\begin{document}

\title{The PRL Stabilized High Resolution Echelle Fiber-fed Spectrograph: Instrument Description \& First Radial Velocity Results}

\author{
  Abhijit Chakraborty\altaffilmark{1},
  Suvrath Mahadevan\altaffilmark{2,3},
  Arpita Roy\altaffilmark{2,3},
  Vaibhav Dixit\altaffilmark{1},
  Eric Harvey Richardson\altaffilmark{4},
  Varun Dongre\altaffilmark{1},
  F. M. Pathan\altaffilmark{1},
  Priyanka Chaturvedi\altaffilmark{1},
  Vishal Shah\altaffilmark{1},
  Girish P. Ubale\altaffilmark{1},
  B.G. Anandarao\altaffilmark{1}
  }

\email{abhijit@prl.res.in}
\altaffiltext{1}{Division of Astronomy,
  Physical Research Laboratory, Navrangpura, Ahmedabad 380015, India}
\altaffiltext{2}{Department of Astronomy and Astrophysics,
  Pennsylvania State University, 525 Davey Laboratory, University
  Park, PA 16802}
\altaffiltext{3}{Center for Exoplanets \& Habitable Worlds,
  Pennsylvania State University, 525 Davey Laboratory, University
  Park, PA 16802}
\altaffiltext{4}{Dept. of Astronomy, Univ. of Victoria, Victoria B.C. Canada; deceased}

\begin{abstract}
We present spectrograph design details and initial radial velocity  results from the PRL optical fiber-fed high-resolution cross-dispersed echelle spectrograph (PARAS), which has recently been commissioned at the Mt Abu 1.2 m telescope, in India. Data obtained as part of the post-commissioning tests with PARAS show velocity precision better than 2~m~s$^{-1}$ over a period of several months on bright RV standard stars. For observations of $\sigma$ Dra we report 1.7~m~s$^{-1}$ precision for a period of seven months and 2.1~m~s$^{-1}$ for HD 9407  over a period of 2 months. PARAS is capable of a single-shot spectral coverage of 3800 - 9500\AA\ at a resolution of $\sim$67,000. The RV results were obtained between 3800\AA\  and 6900\AA\ using simultaneous wavelength calibration with a Thorium-Argon (ThAr) hollow cathode lamp.  The spectrograph is maintained under stable conditions of temperature with a precision of 0.01 - 0.02$^{\circ}$C (rms) at 25.55$^{\circ}$C, and enclosed in a vacuum vessel at pressure of 0.1~$\pm$0.03~mbar. The blaze peak efficiency of the spectrograph between 5000\AA\ and 6500\AA, including the detector, is $\sim 30\%$; and $\sim$ 25\% with the fiber transmission. The total efficiency, including spectrograph, fiber transmission, focal ratio degradation (FRD), and telescope (with 81\% reflectivity) is $\sim$7\% in the same wavelength region on a clear night with good seeing conditions. The stable point spread function (PSF), environmental control, existence of a simultaneous calibration fiber, and availability of observing time make PARAS attractive for a variety of exoplanetary and stellar astrophysics projects. Future plans include testing of octagonal fibers for further scrambling of light, and extensive calibration over the entire wavelength range up to 9500\AA\ using  Thorium-Neon (ThNe) or Uranium-Neon (UNe) spectral lamps. Thus, we demonstrate how such highly stabilized instruments, even on small aperture telescopes, can contribute significantly to the ongoing radial velocity searches for low-mass planets around bright stars.
\end{abstract}



\section{Introduction}
\label{introduction}
\indent The number of known extrasolar planets (exoplanets) has grown rapidly in the last two decades after the discovery of the Jupiter-like companion 51 Peg b \citep{Mayor:1995} using the ELODIE spectrograph \citep{Baranne:1996}. A significant number of the known exoplanets have been discovered or confirmed using the Doppler radial velocity technique \citep[www.exoplanets.org,][]{Wright:2011}. The discovery and characterization of exoplanetary systems via this technique requires high-precision radial velocity time-series spanning months to years, sometimes at high cadence. For instance, the discovery of seven planets around HD 10180 by \citet{Lovis:2011}, was possible only with detailed modeling of the radial velocity (RV) data observed by HARPS \citep{Mayor:2003}, over a period of 6 years at a precision of 1~m~s$^{-1}$ or better. \citet{Tuomi:2013} have recently claimed the existence of  five periodic signals in the RV data on  $\tau$Ceti using a combination of data-sets from three spectrographs: HARPS, KECK-HIRES \citep{Vogt:1994} and AAT(U.C.L Echelle Spectrograph at the Anglo Australian Telescope), spanning a period of 2142 days, and consisting of 4398 RV data points \citep{Tuomi:2013}. A large number of additional RV observations of $\tau$Ceti at high cadence, and at precisions of 1~m~s$^{-1}$ or better, are necessary to determine if these signals correspond to real planets, stellar noise, or artifacts of analysis.

Stabilized fiber-fed spectrographs on small (1-2 m) telescopes are able to fill such a niche and contribute significantly towards the overall understanding of planetary systems. The ability to observe with a dedicated instrument allows the construction of very long baselines whereby fundamental questions, such as the range of planetary system architectures and statistics of planets in the habitable zone, can begin to be addressed. The targeted utility of small telescope instruments lies in the fact that a large number of stars up to 9th magnitude are yet to be monitored with high-resolution spectroscopy; and high cadence observations up to 7th magnitude at a precision of 1-2~m~s$^{-1}$ are similarly lacking. It is only facility-class instruments on small telescopes that can provide both the quality and quantity of data required, without being prohibitively expensive. Indeed, significant contribution has been made in the field of RV exoplanet detection by the CORALIE spectrograph, using the 1.2 m Swiss Euler telescope \citep{Udry:2000,Santos:2003}. Similarly the SOPHIE spectrograph on the 1.9 m telescope in France \citep{Bouchy:2013}  and the recent efforts with the CHIRON spectrograph on the CTIO 1.5 m \citep{Schwab:2012}, are examples of high cadence RV observation efforts on small telescopes, specifically for exoplanet science.

At the Physical Research Laboratory (PRL) in Ahmedabad, India, we have started our own exoplanet program using the technique of precision Doppler RV measurements. We have designed and built an optical fiber fed high-resolution echelle spectrograph (PARAS), which is attached to our 1.2 m telescope at Mt. Abu, India. PARAS is an acronym for the PRL Advanced Radial-velocity Abu-sky Search. The Mt. Abu telescope is located in the western part of India, near the edge of the Thar desert, at an altitude of 1680 meters. The site enjoys good seeing conditions, with median seeing of 1.3" from October to February, and about 2.0" from mid-March to June. Monsoon affects observations between July and mid-September. Typically, we get about 220 cloud free nights in a year, out of which 150 nights may be photometric. The exoplanet project was conceived at PRL with an aim to build a high-resolution optical fiber-fed spectrograph, in a temperature and pressure controlled environment, that would get 80 -- 100 nights in a year on the 1.2m Mt. Abu telescope. Our primary motivation is to target exoplanets around G \& K dwarfs, and to achieve long term stability of 1 - 2~m~s$^{-1}$ on bright stars (up to 7th magnitude) and 5 - 10~m~s$^{-1}$ on fainter targets.

The PARAS spectrograph utilizes the wavelength range between 3800\AA\  -  6900\AA\  for RV measurements, thus maximizing the Q-factor of the data \citep{Bouchy:2001}, and making it possible to achieve 1 - 2~m~s$^{-1}$ on bright targets. The region beyond 7000\AA\ is populated by very bright Argon lines, as well as a large number of atmospheric absorption features, making it less valuable when onbserving FGK stars using a Thorium-Argon simultaneous wavelength calibration source. The optical design concept and the initial engineering test results can be found in \citet{Chakraborty:2010}, and \citet{Chakraborty:2008}. Here we describe the spectrograph, including key elements of the optical design, and report the stability of the instrument, along with new RV results on $\sigma$Dra and HD 9407.


\section{Optical Design of the Spectrograph}

\begin{figure*}[] 
  \centering
  \includegraphics[angle=0, width=1.0\textwidth]{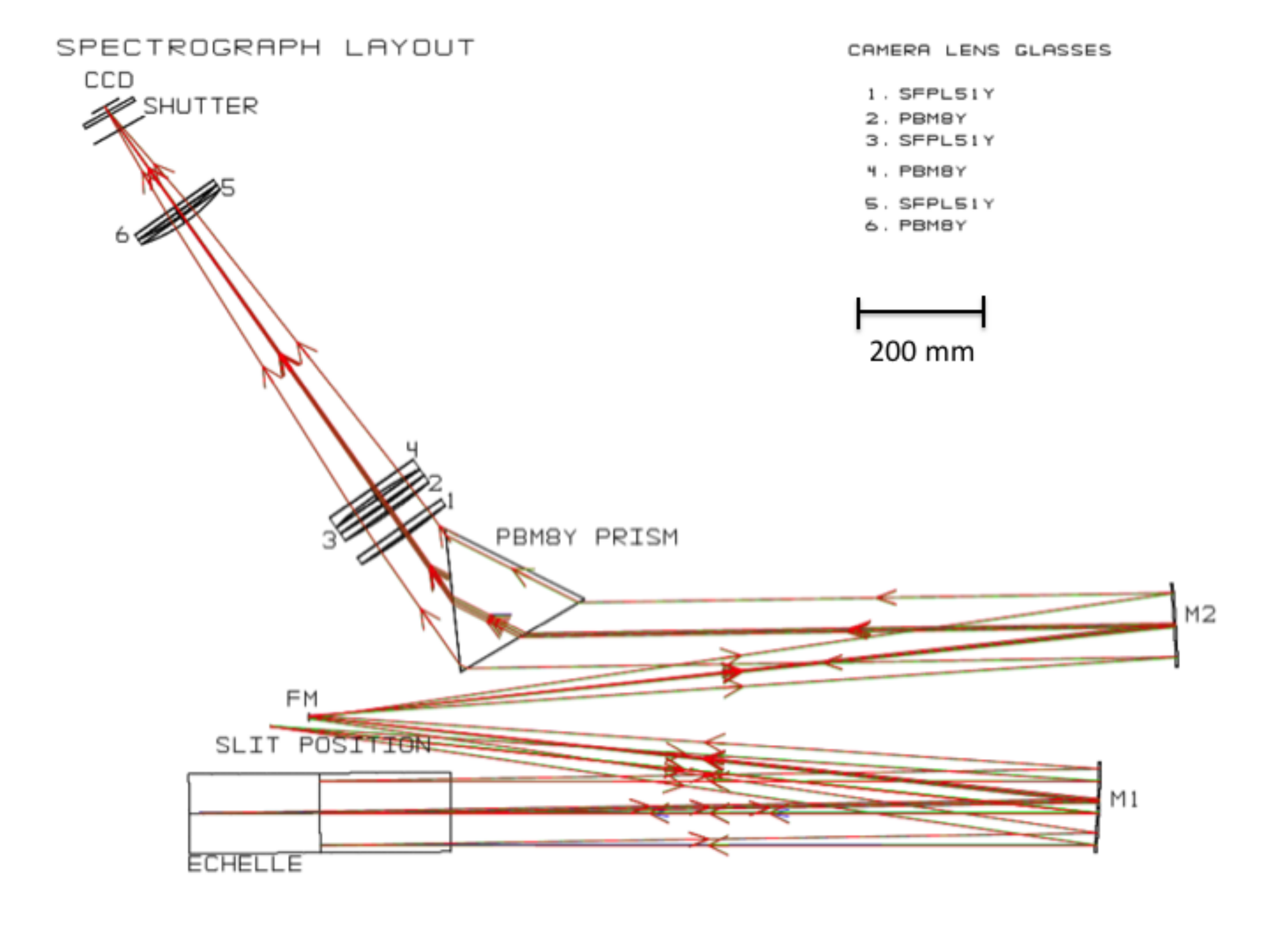}
   \caption{The optical layout of the PARAS spectrograph showing the reflective collimator mirrors, grating, and the glasses that make up the prism and camera optics. Also shown is the scale-bar.}
  \label{fig:opticlayout}
\end{figure*}

A white pupil design was chosen for the PARAS spectrograph in order to enable high RV precision while minimizing issues of scattered light from the ThAr lamp. Figure \ref{fig:opticlayout} shows the optical layout of the spectrograph, and the glasses used in the construction of the prism and the refractive camera. The primary disperser is a 200 $\times$ 400 mm R4 echelle grating with a blaze angle of 75$^{\circ}$ and groove frequency of 31.6 lines/mm replicated on a monolithic Zerodur substrate. This grating, purchased from Richardson Gratings, Newport Corporation, is replicated via a sub-master from the MR160 master grating. The echelle is operated close to the Littrow condition, with an out-of-plane ($\gamma$) angle of 0.45$^{\circ}$. This R4 Echelle, coupled with a large single 4096~$\times$ ~4096 CCD format (62mm x 62mm; pixel size of 15$\mu$m), is sufficient to cover a wavelength range of 3800\AA\ to 9500\AA\ in a single shot, at a resolution R $\sim$ 67,000. The details of the optical ray trace can be found in \citet{Chakraborty:2008}, along with discussions about optimization of the optical design and glass choices for the prism and camera optics.

The total spectrograph efficiency from the slit position to the detector remains greater than 30\% with the PARAS design. We found the Ohara glasses PBM8Y and S-FPL51Y particularly suited to our requirements, owing to their high blue transmission efficiency. The choice of a single large prism cross-disperser with apex angle of 65.6$^{\circ}$ yields inter-order separations of $\sim$~74 pixels at 3800\AA\ , $\sim$~32 pixels at 6900\AA\ , and $\sim$~23~pixels at 9500\AA. This allows two-fiber simultaneous star-ThAr science observations between 3800\AA\ and 6900\AA\ without significant intra-order contamination. The spectra from two fibers are separated by 17 pixels on the detector, and the PSF has a FWHM of about 4 pixels. The details of the individual optical components along with their transmission losses (for lenses) and reflectivity (for mirrors) as manufactured by SESO can be found in \citet{Chakraborty:2010}. The surface micro-roughness of the off-axis parabolic mirrors and the flat mirror were measured and found to be better than 4~nm. This ensures minimal scattering of light within the spectrograph.

Table \ref{table:ZE} shows the optical surface data from Zemax for the spectrograph. The final optical surfaces of the lenses, prism and the mirrors have been re-optimised by SESO using actual measured refractive indices of glass melts. The glass quality used is of homogeneity class H4, limiting the maximum allowed variation in the refractive index in any dimension to below 2$\times$10$^{-6}$. 

At the SESO facility, the individual lenses were first aligned and glued together and then fitted inside the Camera barrel. Before freezing the lens positions inside the barrel the camera lens system was subject to interferometric tests for EE80 (80\% encircled energy diameter test) for spots. The requirement was of less than two pixels (less than 30 microns), however, the test showed EE80 within 10 microns. The mirror mounts were also prepared at the SESO facility, no tip-tilt facility was provided with the mirror mounts but made sure that the reflective face centre is perpendicular to bottom base. The mounts were very simple with a stable broad base and the mirrors were fixed such that their centers were 300mm high from the bottom. It was decided to put the optic axis of the spectrograph 300mm high from the vacuum vessel inside bottom surface which is defined as the optical bench. The echelle and the prism mounts were made by Aditya High Vacuum which also made the vacuum vessel. The echelle surface is oversized by about 6mm on all four sides, and in this narrow strip of 6mm there are no grating rulings. This narrow strip was used to hold the Echelle face-down very precisely and securely thus avoiding any gluing of the echelle from the top. Since the echelle is installed under the littrow condition which is 75 degrees, it rests at an inclined position of 15 degrees (w.r.t. optical bench) which takes away most of the weight of the echelle. We found this configuration very stable particularly for RV measurements. Similarly the Prism is also slightly made oversized by about 10mm and it rests on a triangular shaped mount with a 10mm high step at the edge. We have allowed a gap of 0.5mm between the prism surface and the step, which is cushioned by vacuum compatible teflon sheets.  
 
We aligned the spectrograph inside the vacuum vessel in two steps: 1) A dummy-plate which is identical in shape and size of the bottom plate of the vacuum vessel was made. Next, the Zemax optical layout of the spectrograph was plotted with a size equal to the actual size of the optical components of the spectrograph using a vary large plotter. This optical plot was then spread on the dummy-plate and glued. We then put all the optical components along with their mounts including the CCD dewar on the dummy plate as per the Zemax layout. We then optimised the positions of all the optical components to build the spectrograph on the dummy-plate. Having satisfied with spectrograph performance in terms of resolution and spectral coverage, we marked the position of all the optical components and the CCD on the dummy-plate. 2) Since the dummy-plate was identical in shape and size of the vacuum vessel bottom plate where the spectrograph would be finally built, the final positions of all the spectrograph components from the dummy-plate were easily transferred to vacuum vessel bottom plate within an error of a millimeter, which were easily optimised inside the vacuum vessel. The procedure also helped us to circumvent the difficulty of aligning optical components inside a vessel with restrictions of movements.   

\section{Fiber Feed}
\subsection{Fiber feed at the Spectrograph Interface}
\begin{figure*}[] 
  \centering
  \includegraphics[width=1\textwidth]{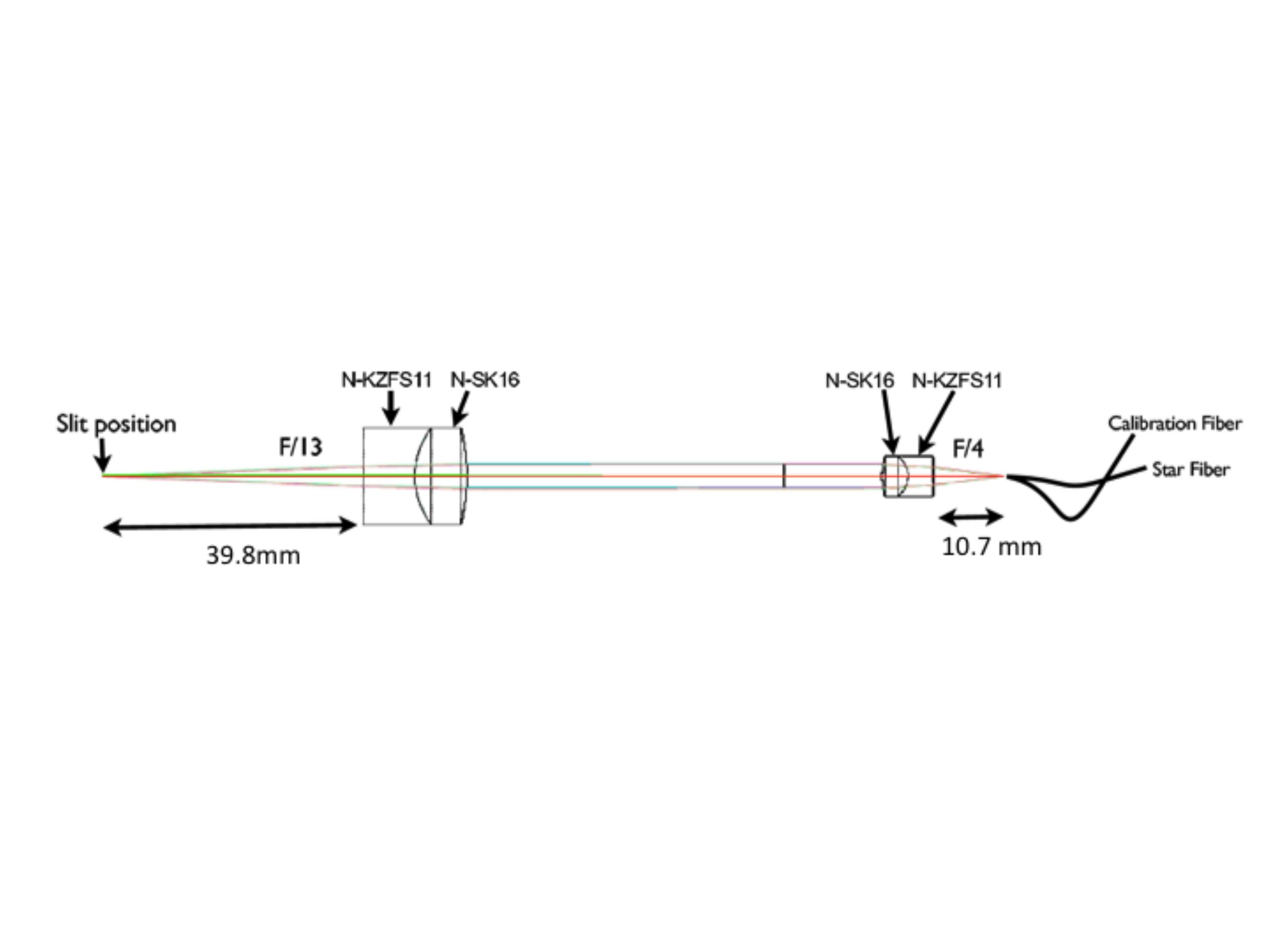}
   \caption{Optical layout showing the projection of fiber tips onto the slit position of the spectrograph. The optics accept F/4 cone angle from the fibers and projects an F/13 beam onto a virtual slit position. The centre to centre separation of the two fibers is 180$\mu$m $\pm3$$\mu$m.}
  \label{fig:fiberfeed}
\end{figure*}

PARAS uses 20 meters of 50$\mu$m diameter circular Polymicro FBP type fibers to couple the telescope to the spectrograph. The star fiber and the calibration fiber at the spectrograph end are part of a fiber bundle (with three dummy fibers) such that these two fiber (50$\mu$m) cores are separated by 180$\pm$3$\mu$m (center to center); the dummy fibers within the bundle keep the star and calibration fibers in their positions respectively with minimal effect of temperature variations. The fiber bundle was made by Polymicro technologies and uses ST connectors on both ends. The 180$\mu$m separation projects to a 17 pixel separation on the detector. There is no other physical slit in the spectrograph beyond the fibers. The F/4 beam coming out of the fiber tips is first collimated by a F/4 achromatic doublet. The fiber tips are then re-imaged onto a virtual slit position using a F/13 achromatic doublet. Figure\ref{fig:fiberfeed} shows the F/4 to F/13 optics. Both the doublets are custom made achromats yielding geometric spot sizes within 20$\mu$m diameter across the spectrograph wavelength range. The projected fiber image at the slit position defines the resolution, thus avoiding slit losses and, most importantly, stabilizing the spectrograph PSF which enables 1~m~s$^{-1}$ precision. The absence of a physical slit also ensures that modal noise issues are not exacerbated by modal filtering.

\subsection{Fiber feed at the Telescope Interface}
Some of the primary issues that we faced while coupling the spectrograph fiber optics to the Mt. Abu telescope were those of telescope flexure and tracking error. The telescope system is aged, has innate tracking accuracies of only 1" - 1.5", and suffers from hysteresis in RA and DEC drives. Due to flexure alone, the image of the star can move in the focal plane by several hundred microns as the telescope tracks an object from east to the west.

\begin{figure*}[] 
  \centering
  \includegraphics[width=0.7\textwidth, angle=0]{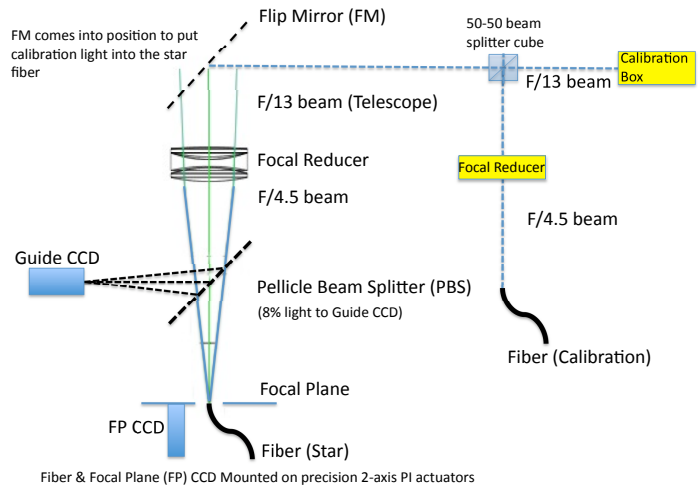}
   \caption{Schematic of the Cassegrain unit (not to scale) showing how starlight and calibration light are fed into the star and calibration fibers respectively.}
 \label{fig:cass}
\end{figure*}

We adopted the following procedure to feed the starlight into the fiber (Fig.\ref{fig:cass}). The 1.2m telescope F/13 beam is converted into F/4.5 using a focal reducer, and the 50$\mu m$ star fiber tip is directly inserted into this F/4.5 beam. The fiber thus sees about 1.9" on the sky. The PSF spot-size at the focal plane is 55$\mu m$ geometric diameter and about 40$\mu m$ rms diameter. 40$\mu m$ is 1.8" on the sky, therefore, the light from the star overfills the fiber. Typical observed star FWHM at the F/4.5 focal plane is around 1.8" when the actual seeing is 1.6" or better (less than 1.6"). While this results in light loss in good seeing conditions, it also provides a more stable PSF at the fiber input. Thus, PSF variation issues are mitigated by damping out seeing and guiding related PSF changes.

Using two precision actuators (X-axis and Y-axis) from Physik Instruments (PI), a mechanical arm is used to move the fiber tip in the telescope focal plane with precisions better than 3$\mu m$. The arm holding the fiber tip is also equipped with a small CCD camera (uncooled Lodestar CCD from Starlight-Xpress). The distance between the CCD centre and the fiber tip is known to high precision (27mm$\pm$ 5$\mu$m). The system is also guided by a Starfish CCD which gathers  8\% of the unfiltered light using a pellicle beam splitter placed after the focal reducer in the F/4.5 beam. The PI actuators first move the CCD camera into the telescope beam; the star focus is checked and the Starfish CCD locks on the guide star. Next, the fiber tip is moved into the position of the CCD Camera and centroids on the star. A schematic of the fiber feed is shown in Figure \ref{fig:cass}.

There is an exposure meter (another Lodestar CCD), mounted at the spectrograph interface, which measures the total unfiltered light coming out of the fiber. This is conveniently achieved by introducing a flip mirror between the last fiber optics lens (F/13 doublet) and the slit position inside the vacuum chamber. The flip mirror directs the starlight out of the vacuum chamber through a small window, and the light is refocussed onto the exposure meter using a commercial achromatic doublet. The exposure meter is used to maximize the star photon counts on the fiber, by moving the PI actuators. Once this is optimized, the mirror is flipped back for the starlight to enter the spectrograph. Typical overhead times; which include telescope pointing, guiding on the star, moving the fiber tip and centering it on the star, and maximizing photon counts, is about 5 - 7 minutes. It is worth mentioning here that during the actual stellar exposure, the guide-star (Starfish) CCD at the cassegrain end of the fiber stores a log of the guiding errors and corrections applied, and continuously updates the image of the star at 1Hz frequency inside a small box of size of 5 $\times$ 5 arcsecs. This procedure gives us a good idea of the sky conditions during the exposure and thereby maintains a good data quality.

\section{CCD}
To enable efficient coverage of the broad bandwidth of the PARAS spectrograph we selected the 4096$\times$4096 E2V CCD231-84-1-D81 device as the detector array. This deep depletion CCD has 15$\mu m$ square pixels and a custom E2V ``astro-broadband" anti-reflection (AR) coating that provides high quantum efficiency (QE) over most of the operating wavelength region of PARAS. The deep depletion device does not exhibit any significant fringing at red wavelengths, making the standard coating sufficient for our purposes. Table \ref{table:QE} shows the measured QE for the PARAS science array at specific wavelengths.

The CCD dewar and cryogenic assembly was built by Infrared Laboratories. To minimize vibrations caused by the closed cycle cooling we employ a low vibration Cryo-Cooler DMX-20 head from Advanced Research Systems Inc., which uses a helium closed cooling system to achieve cooling as well as a vibration amplitude of less than 120 nm at the detector plane. A Lakeshore model 331 temperature controller is used to maintain the CCD at its optimum operating temperature of -115$^{\circ}$C, with a precision of 0.01$^{\circ}$C. The CCD control and data acquisition are accomplished with a SDSU Gen-III controller and associated electronics. A computer running Windows 7 and Owl software (supplied with SDSU Controller) is used to acquire and display the data. The CCD exhibits dark noise of $\sim 0.02$ electrons/pixel/hour at the standard operating temperature, and the slow-readout mode (100KHz; 153secs total readout time) with the controller yields a read noise as low as 2.1 electrons as tested in the laboratory. Performance at the observatory is worse due to  grounding and RF noise issues at the mountain top. The RF noise is picked up by the controller to the CCD cable, and measured read noise in the observatory is approximately 4 electrons.

\section{Environmental Control: Pressure \& Temperature}

\subsection{Vacuum Vessel Stability \& Performance}
\begin{figure*}[] 
  \centering
   \includegraphics[width=0.9\textwidth]{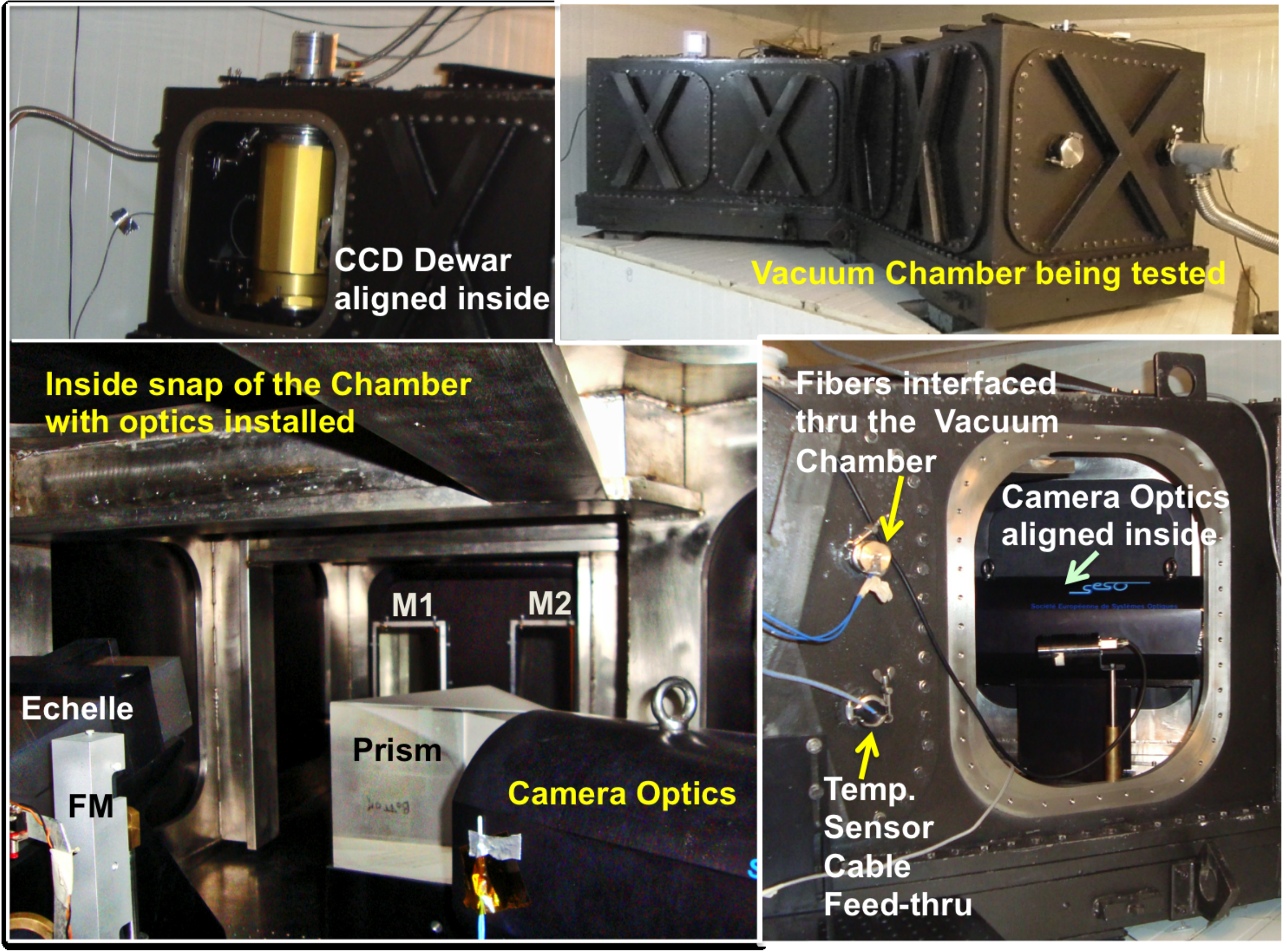}
      \vspace{10pt}
       \caption{PARAS in the vacuum chamber. {\it Top left:} The CCD dewar aligned inside the vacuum chamber. {\it Top right:} The vacuum vessel being installed inside the inner room. {\it Bottom left:} Some of the optical components seen in the chamber where M1 \& M2 are the off-axis parabolic mirrors and FM is the Fold mirror as in figure 1. {\it Bottom right:} The optical fiber interface with the vacuum chamber and part of the camera lens system.}
   \label{fig:vacuumvessel}
\end{figure*}

The PARAS vacuum enclosure is made of Stainless Steel 304L and is Rectangular in shape. It has large rectangular port openings for easy access to optical components placed inside the vessel.  The ports are covered using aluminum flanges.  Both the materials are vacuum compatible and can hold vacuum with vacuum pumps isolated, due to low out-gassing properties. The shape of the vessel was optimally chosen to enclose the entire optical path from the slit position to the CCD dewar. The long axis of the vessel is $\sim$2 m, while the short axis is $\sim$1.6 m.  The height of the vessel is 0.7 m and is designed to partly enclose the CCD dewar, leaving the helium line connections for cryo-cooling outside the vacuum chamber. This is arranged using flexible stainless steel bellows. See Figure \ref{fig:vacuumvessel}.

The CCD dewar is mounted on a linear stage which can be moved with a precision of 20$\mu$m to adjust the focus. This is sufficient, considering that the depth of focus is 60-70$\mu$m. The precision stage is moved using a handle outside the vacuum chamber, which is connected to the stage through a vacuum feed-through. The base plate of the vacuum chamber acts as the optical bench and has a surface flatness of 20$\mu$m, with the optical axis set 300 mm above the optical bench. The base plate and the port flanges are strengthened by external ribs, and the vessel walls are strengthened by internal ribs to minimize deformation under vacuum. Aditya High Vacuum, Ahmedabad, India manufactured the vacuum chamber. Details of the analysis performed with ANSYS are mentioned in \citet{Chakraborty:2010}.

The evacuation of the vessel is performed using a dry vacuum pump (Alcatel make, multi-roots type). The pump has a vacuum capacity of 0.05 mbar over this large volume. The vacuum is measured using an absolute pressure measurement gauge using the capacitance manometer principle.  Leak issues in early 2011 at the CCD linear-stage vacuum feed-through and the bellows on top of the CCD dewar have now been resolved to an acceptable leak rate. Typical absolute radial velocity drift is expected to be about 1~m~s$^{-1}$ per 0.01 mbar pressure change \citep{Pepe:2002}. Therefore, over the course of the night we did not want the absolute drift to be more than 20~m~s$^{-1}$ due to pressure change alone. It is worth noting here that the simultaneous reference technique requires that the air or medium between the optical trains coming out of the two fibers remain as similar as possible. The vacuum assists this by preventing formation of any air turbulence or eddies within the spectrograph. Typically, we have pressure increases of $<$0.06mbar during a night of observations (0.07 mbar to 0.13 mbar in about 15 hours).

The star fiber and the calibration fiber are channeled into the vacuum vessel using a special fiber feed-through, where the fiber teflon sleeve is symmetrically compressed to a diameter of 90$\mu$m, which only slightly squeezes the fiber. A helium leak test of the connectors prove that the leak rate is on the order of 10$^{-9}$ mbar-liter/sec through the teflon sleeve, and this leak is less than the overall leak rate from the entire vessel which is 10$^{-8}$ mbar-litre/sec. The electrical connections are led into the vacuum chamber through vacuum compatible multi-pin feed-through. Figure \ref{fig:vacuumvessel} shows several views of the vacuum vessel with optical components inside.

\subsection{Temperature Control \& Performance}
The temperature control environment for the spectrograph mainly consists of two concentric outer and inner rooms. These are made of highly insulated 60mm thick PUF (Polyurethane Foam) material. The outer room is 14 $\times$ 11.5 feet, while the inner room is 11 $\times$ 9 feet. The outer room temperature control is held at 23.5 $\pm$0.8$^{\circ}$C, using standard off-the-shelf commercially available PID control of simultaneous air-conditioning (cooling \& heating). The inner room has an in-built temperature control system which primarily uses heaters. The temperature is controlled by holding it 2$^{\circ}$C above the outer chamber temperature set-point. We have used 10 panels of 30 Watt heaters in the inner room. These heaters are fixed on the walls of the inner chamber surrounding the vacuum vessel housing the spectrograph. The electronic unit is based on the micro controller 89S51. We have used Pt100 temperature sensors (Platinum resistive sensor), which are known for their high stability (0.005\% over a period of a year and accuracy of 0.01$^{\circ}$C \footnote{ http://www.omega.com/prodinfo/rtd.html}). A detailed discussion on the temperature control system is provided by Dongre et al. (2013, PRL technical note\footnote{PDF available for download at http://www.prl.res.in/$\sim$abhijit}). As mentioned in Dongre et al. (2013) we found a 0.5$^{\circ}$C ambient temperature difference between the inner room and the vacuum vessel, primarily due to heat leakages from the base of the vessel to the concrete pillar, in spite of 4 inch thick insulated materials. This leak was subsequently mitigated using extra calibrated heat pads and air circulating fans around the vessel base. Further minimization of heat leak may be necessary for more precise temperature control.

\begin{figure*}[ht] 
   \centering
   \includegraphics[width=0.8\textwidth, angle=0]{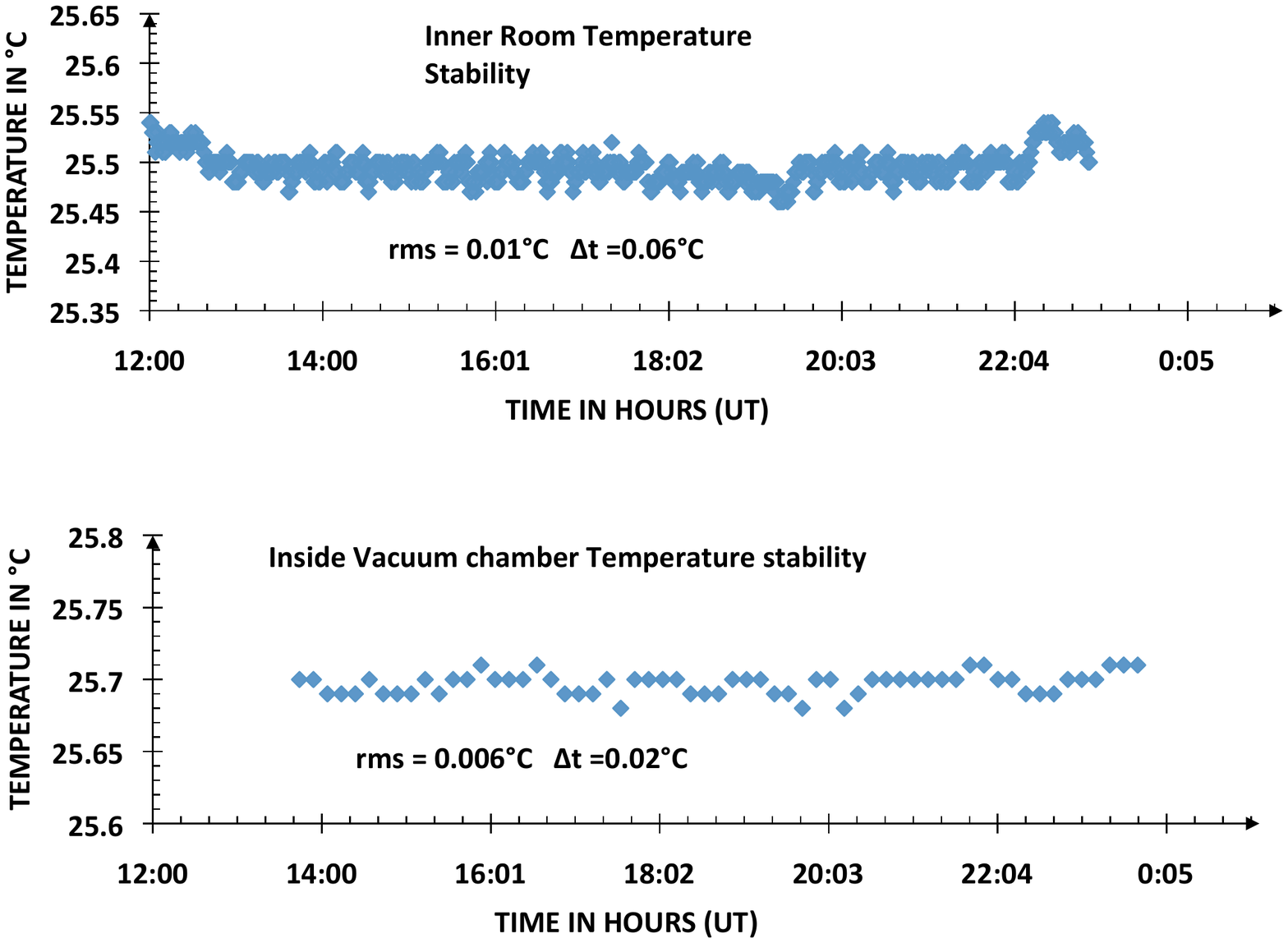}
      \vspace{10pt}
       \caption{Temperature variation over a period of 12 hours during the night of observations. {\it Top}: Temperature variation in the inner room at around 25.5$^{\circ}$C. {\it Bottom}: Corresponding temperature variation inside the vacuum vessel. The sensor is placed on the camera optics metal body-tube.}
   \label{fig:parasstability1}
\end{figure*}

Figure \ref{fig:parasstability1} shows the stability in the inner room; an independent temperature sensor attached near the Camera Lens Assembly tube, inside the vacuum vessel, shows the temperature stability inside the vessel. Note that there is an DC offset of 0.2$^{\circ}$C between the two sensors. Thus, maximum temperature variations inside the vacuum vessel during the night are $\sim$0.02$^{\circ}$C and in the inner room is $\sim$0.06$^{\circ}$C. The thermal inertia of the vessel gives a better stability. In future, we plan to further improve the control system to 16bit resolution with multiple sensors and control the maximum change with a resolution of 0.001$^{\circ}$C and accuracy of 0.01$^{\circ}$C. This will help bring down the temperature variation, ${\Delta}$T, closer to 0.01$^{\circ}$C in the inner room and less than 0.01$^{\circ}$C inside the vacuum vessel.

\section {The Spectrograph Format, Resolution, \& Efficiency}

\begin{figure*}[] 
   \centering
      \includegraphics[width=1.0\textwidth]{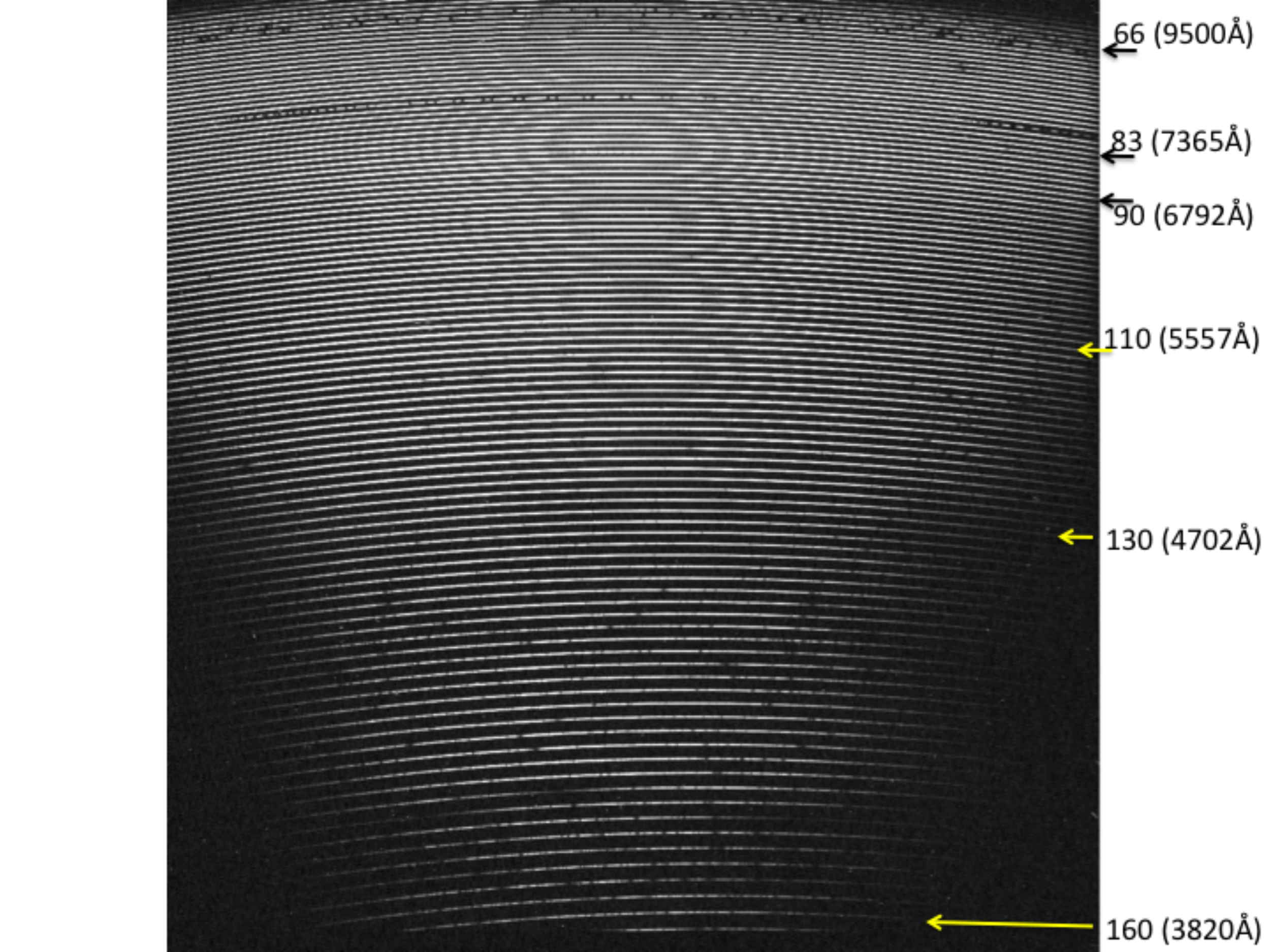}
    \caption{Full-frame raw image of a star (47 UMa), with single fiber illumination. Some of the echelle orders and the corresponding central wavelength are shown with an arrow. Thus the useful orders that can be extracted are from order 160 to 66, the orders below 66 are too close and may suffer from inter order contaminations.}
    \vspace{10pt}
   \label{fig:fullstar}
\end{figure*}

The spectrograph format is designed to acquire a wavelength coverage from 3800\AA\ to about 9500\AA, in a single shot (exposure) on the $4096 \times 4096$ pixels CCD detector. The use of a prism cross-disperser results in the spectral format having decreasing order separation with increasing wavelength (dispersion is approximately inversely proportional to the cube of the wavelength \citep{Schroeder:2000}). Thus, the separation between orders 160 and 159 in the blue is 74 pixels (around 3800\AA ), and the separation between orders 65 and 66 in the red is 23 pixels (around 9500\AA ). Figure \ref{fig:fullstar} illustrates a full spectrum of the star 47 UMa with single fiber illumination. However, the general operating mode uses two fiber illumination for simultaneous wavelength calibration, where the corresponding spectral orders are separated by at least 17 pixels on the detector.

\begin{figure*}[] 
   \centering
   \includegraphics[width=0.85\textwidth]{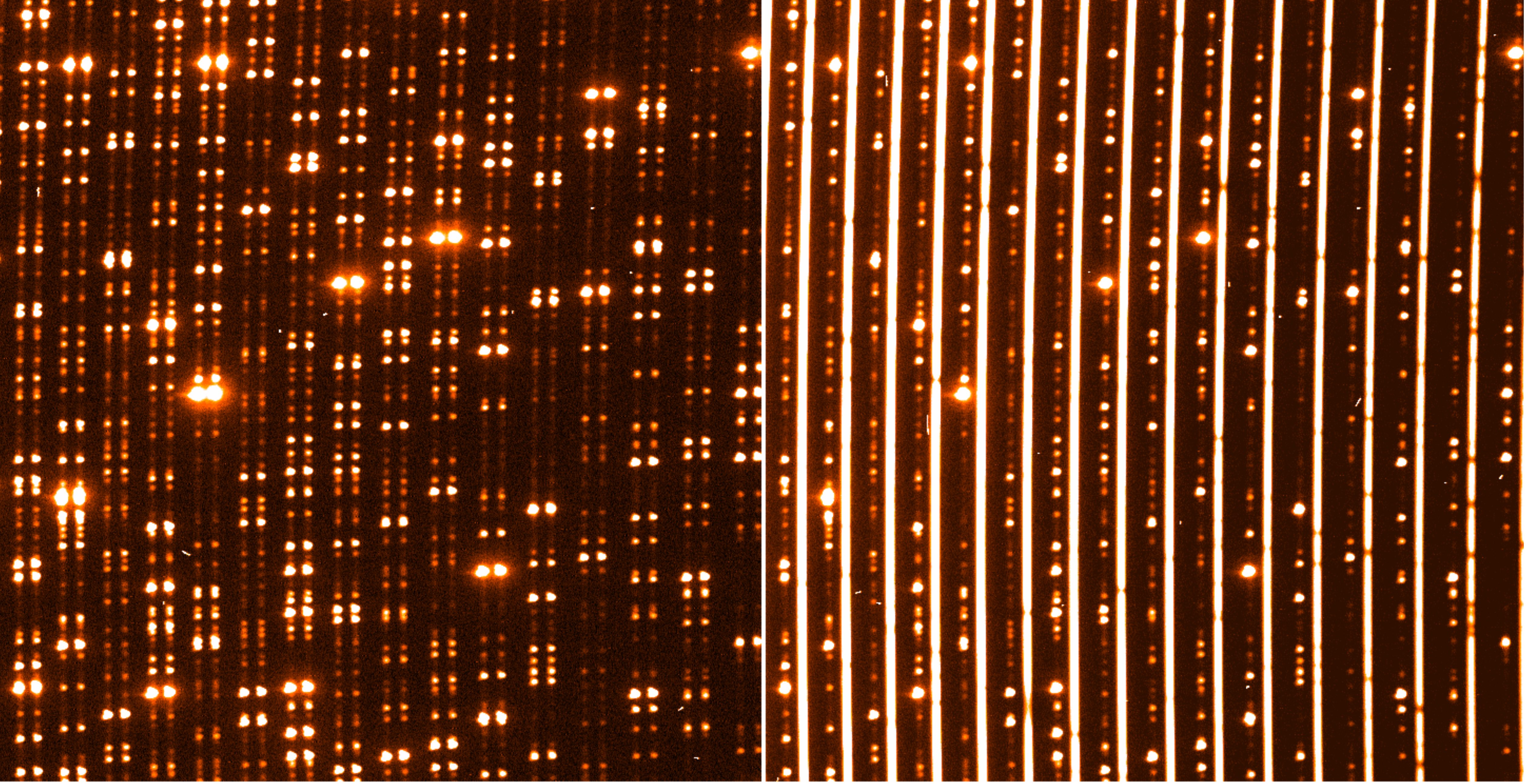}
    \caption{Portion of raw image for {\bf (left)} ThAr-ThAr exposure, and {\bf (right)} stellar exposure with simultaneous ThAr calibration.}
   \label{fig:frames}
\end{figure*}

\begin{figure*}[] 
  \centering
  \includegraphics[width=0.7\textwidth]{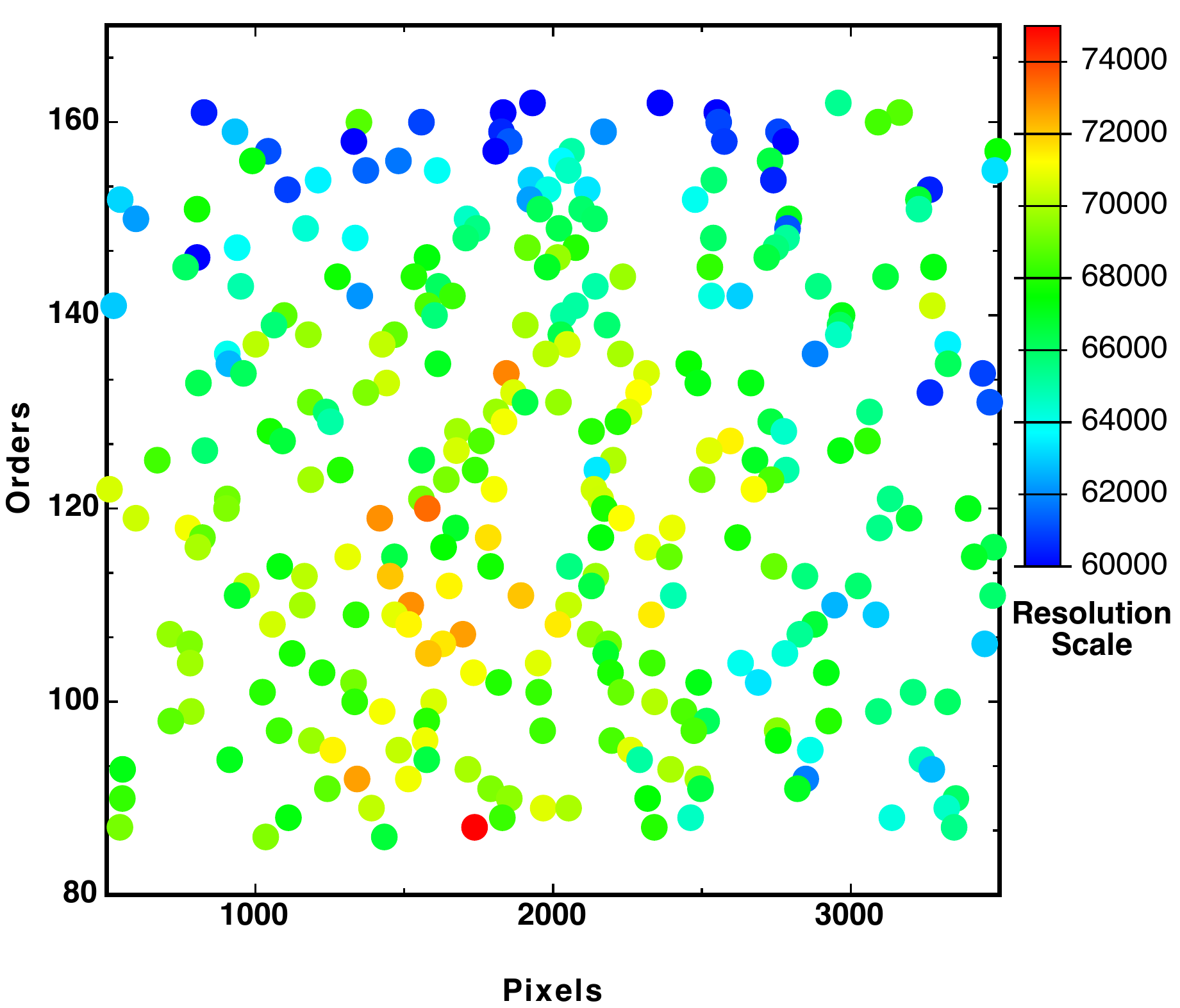}
       \caption{Measured spectral resolution of PARAS between 3800\AA\  \& 7000\AA. The x-axis is in pixels and represents the actual
   CCD scale; the left y-axis is in orders and represents the portion of the chip underlying extracted spectra.
   For illustrative purposes, the orders are shown only in one dimension (removing echelle order curvature).
   The resolution scale is shown on the right side. From each order some 5 to 6 clearly resolved Thorium lines have been used
   to measure the spectrograph resolution. Note that on the shorter wavelength side (higher orders) the measured resolution is slightly
   lower at about 60000 and on the longer wavelengths up to 7000\AA\ it is about 75000. The median resolution of the spectrograph between 3800\AA\ and 7000\AA\ is
   about 67000.
   }
  \label{fig:res}
\end{figure*}

Two fiber illumination is only possible down to order 88 (6900\AA ), where the inter-order separation increases to 32 pixels. This spectral region (3800\AA\ to 6900\AA ) is sufficient for precision radial velocity measurements on F, G, K stars \citep{Bouchy:2001}. Figure \ref{fig:frames} shows the region around 5300\AA\ for frames taken during two fiber illumination on the detector. The spectral resolution was determined between orders 160 and 87 using ThAr lines. The FWHM of the ThAr emission lines was found to be between 3.7 - 4.3 pixels. The median FWHM of the PSF is 4 pixels. Figure \ref{fig:res} shows the spectrograph resolution chart between 3800\AA\  (order 160) \&  7000\AA\ (order 87) with a median resolution of 67,000. Because of the presence of very bright Argon lines in the longer wavelengths, it was difficult to measure the resolution at regular intervals beyond 7000\AA. We note that at about 8500\AA\  the resolution is found to be 65000, and at 9000\AA\ about 61000. Thus as shown in figure 8 the resolving power in general increases from 3800\AA\ to 7000\AA. The resolving power however slightly decreases beyond 7000\AA, since the spectrograph is not fully optimized for this region. A detailed resolution map in these redder wavelengths will be possible in the near future using ThNe or UNe spectral lamps.

The Mt. Abu telescope reflectivity is measured periodically using a white-light reflectometer. The telescope mirror is wet cleaned (every 45 days) using distilled water and alcohol, and typical primary mirror reflectivity is maintained at 80 - 90\%. There are, however, occasions when the primary mirror reflectivity has fallen to below 60\% due to prolonged wet conditions after the end of the monsoon season, or due to a sudden dust storm. The telescope primary mirror is aluminized once in 18 months, or when the reflectivity does not improve to 80\% even after wet cleaning. We have measured the total efficiency of the spectrograph at both 90\% and 60\% telescope primary mirror reflectivity conditions.

\begin{figure*}[] 
   \centering
  \includegraphics[width=0.7\textwidth, angle=270]{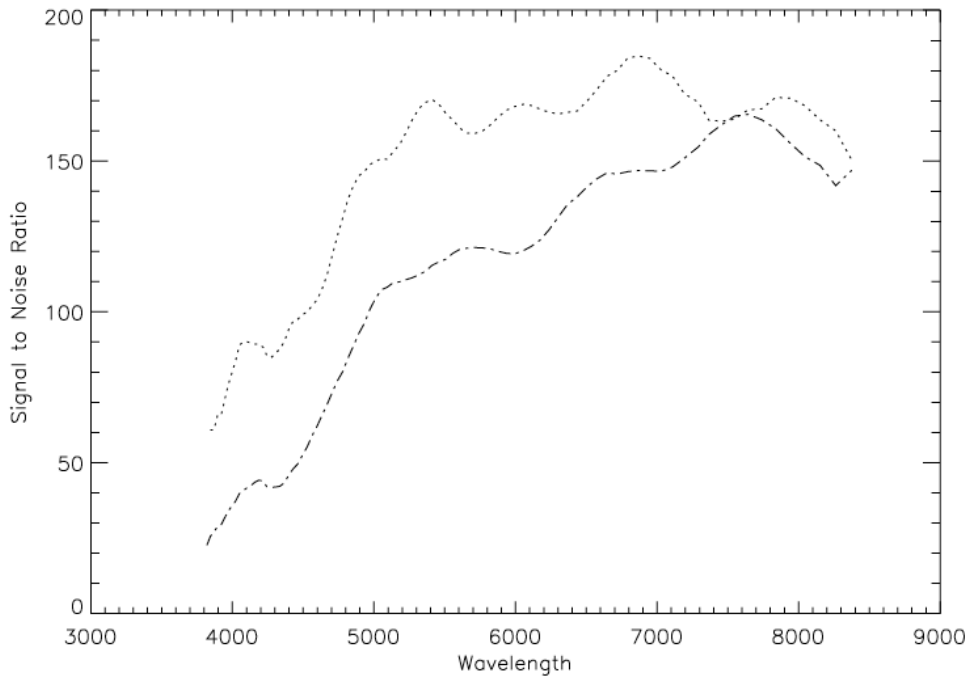}
   \caption{Signal to noise ratio at various wavelengths for the fifth magnitude star 47 UMa, with an exposure time of 10 minutes. The telescope primary mirror reflectivity varies between 90\% (dotted) and 50-60\% (dot-dashed). The sky conditions were similar for both observations, with seeing of 1.8-2.0". The dip at 4200\AA\  is due to a similar dip in the optical fiber transmission, the other dips and bumps seen in the total efficiency curves (the dotted and dash-dotted curves) are most probably due to the variation in the Mt. Abu sky transmission, further investigation in the source of such variation may be needed. The same effect is also seen in figure 10}
   \label{fig:snr}
\end{figure*}

\begin{figure*}[] 
   \centering
  \includegraphics[width=0.7\textwidth, angle=270]{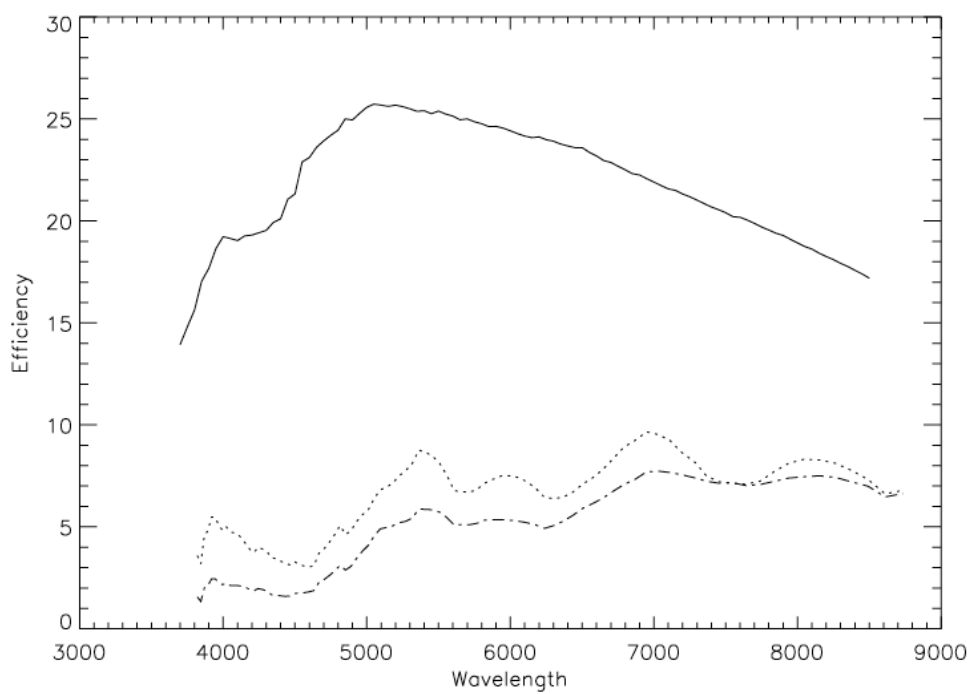}
    \caption{Efficiency of the PARAS spectrograph. The solid line shows the estimated efficiency of the spectrograph, which includes the measured Q.E. of the CCD detector, reflectivity and the transmissivity of the mirrors, blaze peak efficiency of each echelle order, and losses associated with the camera lenses, prism, and optical fibers. The dotted line represents the total efficiency of the spectrograph including the telescope -- this implies guiding losses and a primary mirror reflectivity of 90\%. The dash-dotted line represents the total efficiency of the spectrograph including the telescope, but with the telescope primary mirror reflectivity between 50 - 60\%.}
   \label{fig:efficiency}
\end{figure*}

\begin{figure*}[] 
   \centering
   \includegraphics[width=0.55\textwidth, angle=90]{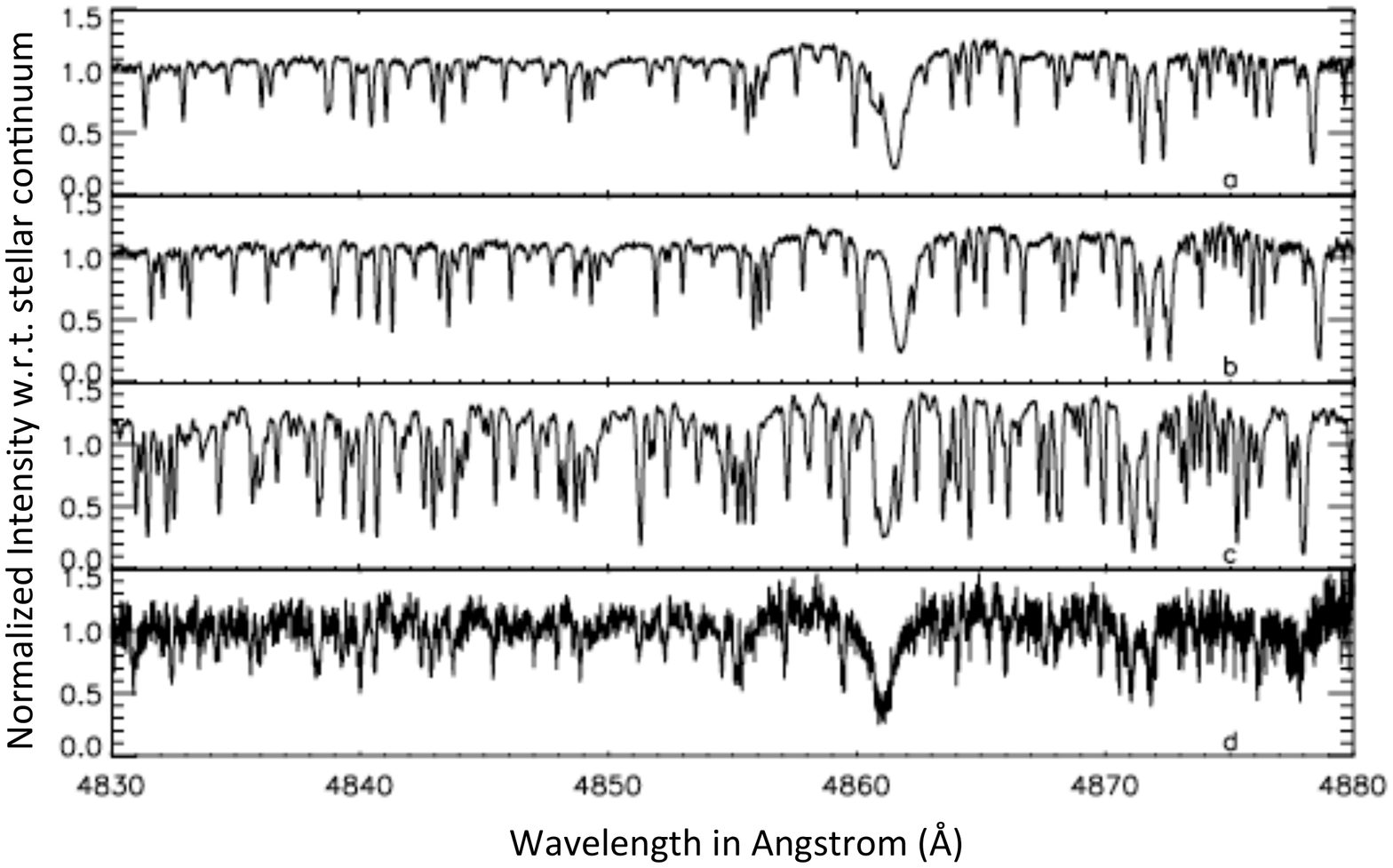}
   \includegraphics[width=0.55\textwidth, angle=90]{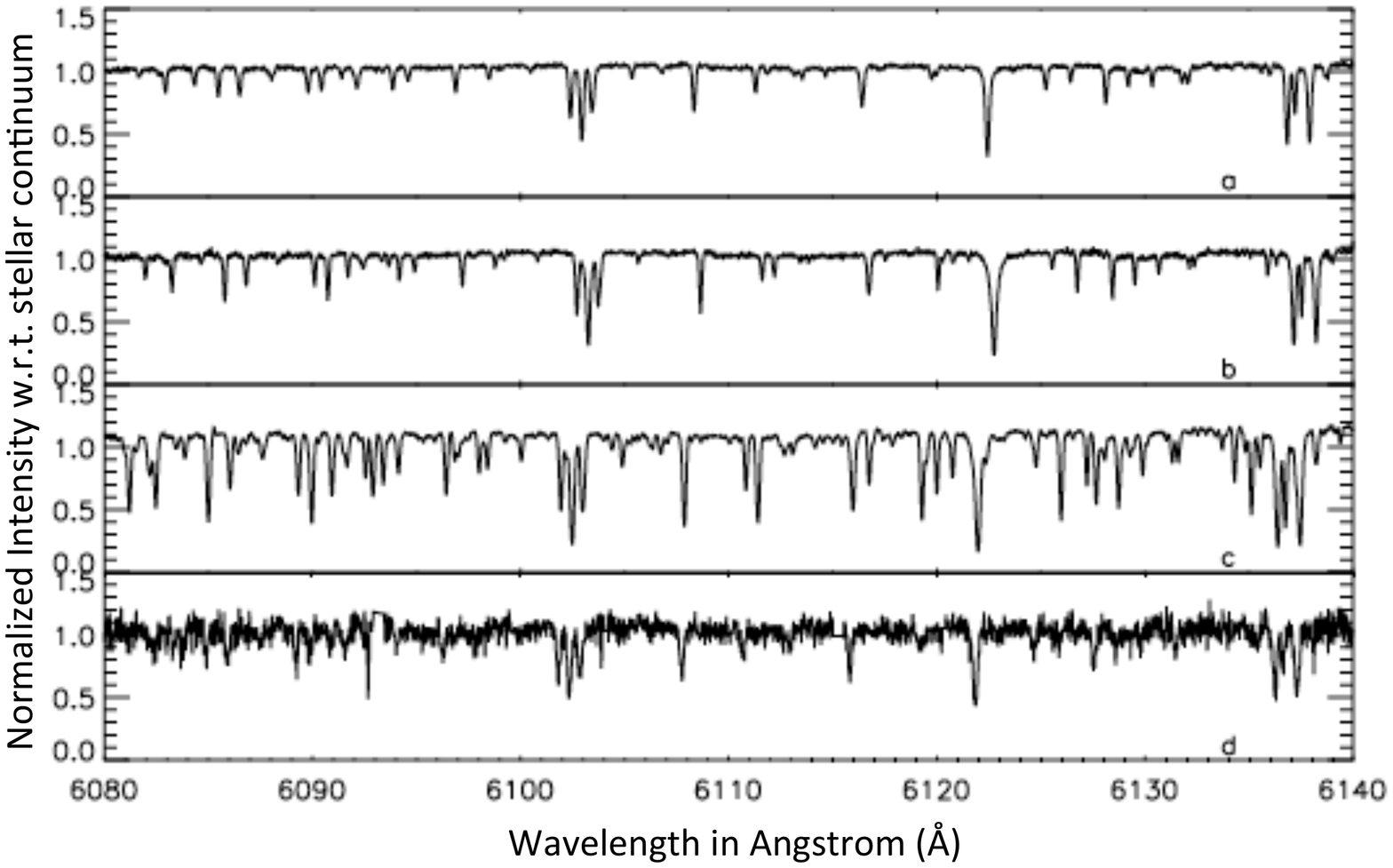}
    \caption{Illustrative section of extracted spectra. The stars shown are 47 UMa (G1V, a),
     Sig Dra (K0V, b), HD137759 (K2III, c) and HD 185374 (G0 IV, d)}
   \label{fig:3stars}
\end{figure*}

Figure \ref{fig:snr} shows the signal-to-noise ratio (S/N) of the star 47 UMa in a 10 minute exposure under 90\% and $<$60\% primary mirror reflectivity. In both cases the secondary mirror had little effect, as it maintained higher reflectivity levels of 85 - 92\%. Figure \ref{fig:efficiency} shows the total spectrograph efficiency under similar conditions. Thus, under 90\% primary and secondary mirror reflectivity and at 1.8" seeing or better, the total efficiency of the spectrograph is $\sim$7\% between 5000\AA\ and 6500\AA. Figure \ref{fig:3stars} shows sample spectra of four stars of different spectral types, over small wavelength regions in both the blue and red. Typical exposure times on these sources were 10 minutes, except in the case of HD 185374 where it was 15 minutes.

\section{Data Analysis Pipeline}

\subsection{Reduction Pipeline}
The data reduction pipeline, written in the Interactive Data Language (IDL), is a fully automated, minimally subjective, highly reproducible, and robust set of image processing algorithms that allows batch-processing of large amounts of data. It requires user initiation but not interaction, unless troubleshooting becomes necessary due unexpected external factors. At the core of the pipeline is the REDUCE data analysis package for conventional and cross-dispersed echelle spectra developed by \citep{Piskunov:2002}, which was adapted and evolved to meet the specific needs of the PARAS spectrograph.

The pipeline uses sorted raw calibration and science data in FITS format to produce optimally extracted spectra of science targets. Composite calibration frames are created initially for each night. A master bias is made by co-adding all nightly bias frames in two groups and then combining them after a comparison for bias shifts, read noise effects, and outliers. This master bias frame is subsequently subtracted from all non-bias images. A master flat is similarly created by co-adding all available flats. Echelle order curvature and locations are determined empirically from a high quality flat field image in order to extract spectra with negligible continuum. REDUCE employs a two-dimensional clustering algorithm for this purpose, which selects potential spectral order pixels and tests the level of clustering before merging and rejecting partial orders, ultimately fitting the merged clusters. Since we have a large number of spectral orders, we enhance the performance of this routine by using a special mask that shrouds low signal corners of the image in addition to defining bad pixels. The order tracing routine requires user interaction for clusters that are not obviously meant to be merged or rejected. Since PARAS is a very stable instrument, in lieu of this interactive step, we have a master order trace available. User intervention is only necessary when something goes wrong and the shift between nights exceeds $\pm$3 pixels (typically the nightly shifts are within 1 pixel).

REDUCE uses a sophisticated decomposition routine to normalize flat field images, obtain order shape functions, estimate scattered light in inter-order gaps, and extract spectral orders. This involves a division of the image into swaths, whereby the spectrum (illumination profile along the order) is separated from the spatial profile for each column (relative illumination profile perpendicular to echelle dispersion). Starting with an initial guess for the spectrum, the spatial profile is fit using an empirically determined mean spatial profile, with weights based on a noise model. Iteration ceases when the change in the deduced spectrum becomes small, producing the best possible slit function and spectrum, under the assumption that the large scale structure in the image is spectral and that the order trace is valid. A major advantage of this method is that by assuming a smooth profile within an image swath, the influence of cosmic rays, bad pixels, and noise is minimized. We further assist the routine by providing known CCD defects in a bad pixel map. It is worth mentioning here that the science-grade CCD has only one bad column towards the edge of the array.

It is necessary to estimate the scattered light background for subtraction from both normalized flat field images and spectra, although the background beneath an order cannot be directly measured. Instead, an interpolation of the background between orders after the isolation of inter-order noise with the decomposition method, provides a good estimate. Sky emission is included in this estimate but the routine is not equipped to handle ghosts or very bright emission lines. The careful instrument design of PARAS ensures that scattered light removal has a minimal effect and that ghosts are not a major concern. There are no ghosts in the spectrograph even in the presence of strong Argon lines beyond 7000\AA. Although PARAS has a filter that sharply depletes light of wavelengths greater than 7000\AA, we find that with a prism cross-disperser, the high mutual proximity of the redder orders requires that the bleeding of bright argon lines into star orders be cured diligently. Even a simple linear column-by-column interpolation routine for scattered light subtraction improves pipeline performance in places where bright argon lines are incompletely removed; but our preferred method is to use a bleeding map, linearly scaled by a global parameter describing variations in lamp brightness \citep[similar to][]{Lovis:2007}.

Flat fielding is necessary to remove pixel-to-pixel quantum efficiency variations, but the master flat must first be first normalized in order to minimize noise contributions and avoid magnifying low-signal regions of the image. Especially for a fiber-fed spectrograph like PARAS, where the flat field images have a spatial profile very similar to science exposures, regions at all signal levels must be carefully weighted. Inter-order regions are set to unity, and the order is divided into swaths before decomposition. The spectral profiles are spliced together to form the shape or blaze function for each order, while the spatial profiles for adjacent swaths are interpolated or extrapolated and subsequently scaled by the blaze to normalize each column segment in the master flat. Care is taken to choose the number of swaths so the spatial profile is oversampled, while keeping the swath width large enough that changes in the point-spread-function along a spectral order can be tracked.

Finally, REDUCE optimally extracts spectra using the same swath-based decomposition method described above. The science image is bias subtracted, divided by the normalized flat-field image, decomposed for scattered light correction, and decomposed again for optimized spectral extraction. Continuum normalization is achieved by dividing out the order shape function previously determined.

\subsection{Analysis Pipeline}

The spectra in pixel-order space must be wavelength calibrated using a template of suitable thorium lines. To create an approximate solution for the new instrument, a cursory line identification was initially performed to predict a two-dimensional wavelength function across the chip. A stabilized spectrograph like PARAS it is possible to have a blueprint wavelength solution of the simultaneous ThAr lamp spectra available to calibrate new data for all time. A new wavelength solution blueprint would only be necessary to create if for some reason the fibers break and are needed to be replaced. Pixel shifts are cross-correlated out and subsequently, the redetermination of emission line centers as indicated by a Gaussian fit mutually produce polynomial coefficients to precisely define the graduation of wavelength across the order. A complete thorium line list for the PARAS spectral range is utilized (similar to the SOPHIE line list at {\it www.obs-hp.fr}). A deviation threshold removes outliers and prevents noise from being mistaken as emission lines. The user can change both the level of error acceptable and the minimum number of lines used to define an order.

\begin{figure*}[ht] 
   \centering
   \includegraphics[width=0.7\textwidth]{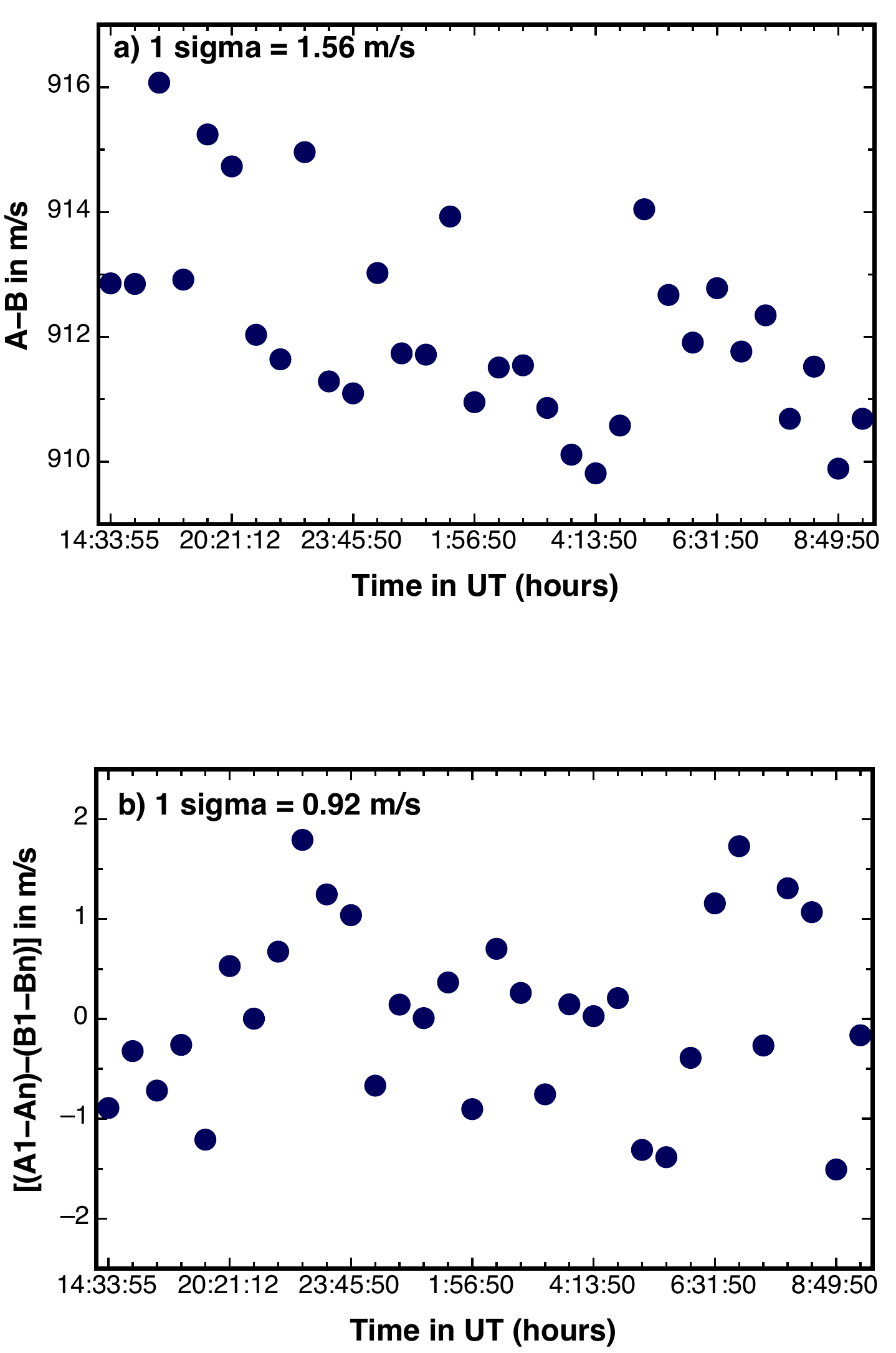}
      \vspace{10pt}
       \caption{{\it Top:} The ``A-B" fiber drift differences over time with 1-$\sigma$ equal to 1.56~m~s$^{-1}$, which is a measure of the spectrograph stability over a night. The x-axis shows time in hours. {\it Bottom:} Precision with which fibers A \& B track each other. (A$_1$-A$_n$) - (B$_1$-B$_n$) between time T$_1$ and T$_n$ over an absolute fiber drift of 160~m~s$^{-1}$ during the night. Here A1, An, B1 \& Bn are the measured absolute drifts of fibers A and B respectively, and $n = 2,3,4..... 34$ (there are 34 measurements in total). This is a better measure of the stability of the instrument, and yields a 1-$\sigma$ value of 0.92~m~s$^{-1}$.}
   \label{fig:parasstability2}
\end{figure*}

In order to track instrument stability, we create a binary mask of reliable sharp thorium lines and use this in conjunction with periodic calibration lamp exposures throughout the night. The cross correlation function (CCF) is calculated for a shifting thorium mask against each spectral order, and the CCFs are summed in order to fit a gaussian peak to the true drift value of the image. For simultaneous thorium exposures on stellar spectra, this also yields the temporal instrument drift correction. Figure \ref{fig:parasstability2} shows the spectrograph stability using simultaneous thorium exposures on both the fibers A \& B. If the internal optical and mechanical parts of the spectrograph are stable, and see the same minute changes in temperature and pressure, then the thorium spectra from fibers A \& B should drift identically. Thus the difference between A \& B thorium spectra in terms of RV at time T should be stable. Typical observed 1-$\sigma$ scatter during a night of observations are 1.5~m~s$^{-1}$ or less. This is shown in Figure \ref{fig:parasstability2} ({\it top}). Spectrograph stability can also be defined as the amount of absolute drift of each fiber, A or B, between time T$_1$ and T$_n$. Consequently, the difference in absolute drifts in Fiber A and B should be (A$_1$-A$_n$) $\sim$ (B$_1$-B$_n$) between time T$_1$ and T$_n$. The 1-$\sigma$ scatter for this measurement is less than 1~m~s$^{-1}$. This is shown in Figure \ref{fig:parasstability2} ({\it bottom}).

Radial velocities are derived by cross correlating target spectra with a suitable numerical stellar template mask based on spectral type (with default G2), created purposely from high signal-to-noise or synthetic data \citep{Baranne:1996}. Mask lines are centered around bright absorption peaks, with a width of 3~km~s$^{-1}$ and a depth set by the star \citep{Pepe:2002}. We begin the algorithm for precision RVs by providing an estimate for the radial velocity of the star, to which the barycentric correction is applied. This centers the region of scrutiny, the range of which is used to recreate the mask at regular intervals to produce a CCF for each order. The maximum number of useable order CCFs (with no saturated regions) are then summed and fit with a Gaussian to obtain the radial velocity of the observation. Instead of using the wavelength calibration of each simultaneous ThAr spectrum, which is influenced by small spectral variations, we use the thorium mask produced drift values to correct for any small instrument instabilities.

\begin{figure*}[htbp] 
   \centering
   \includegraphics[width=0.75\textwidth]{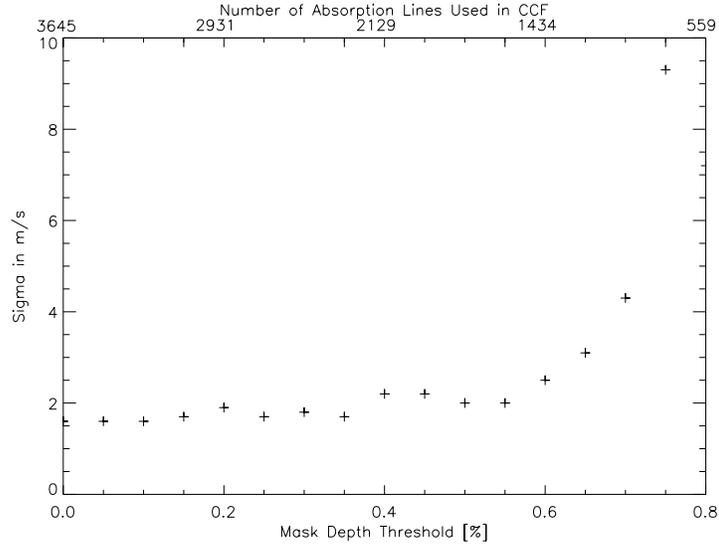}
    \caption{Effect of thresholding line depth of the numerical stellar template mask during radial velocity calculation on the star
    $\sigma$Dra. Mask depth threshold [\%] refers to the minimum percentage depth of lines used for the cross correlation; thus,
    0.4 signifies that all lines with depth greater than 40\% are retained. For low level cuts like 0.2 or 0.4, only the
    shallow lines are discarded, and many lines are used for the cross-correlation, leading to high RV precision. Higher level
    cuts like 0.7 cause most lines to be discarded, leaving only the deepest lines. While deep lines are good for precision,
    the very small total number of lines used in this case hurts precision. It is also interesting to note that there is not a
    loss of precision with mask cut beyond 0.35 or 35\% mask cut, at that point we reach the overall precision of 1.7~m~s$^{-1}$. This is
    most likely to be the star's intrinsic jitter.}
   \label{fig:maskcut}
\end{figure*}

On testing the effect of using all available mask lines on RV standard star $\sigma$ Dra, we find that using more lines leads to better RVs as expected, but also that a noise floor of about 1.7~m~s$^{-1}$ is reached with lines deeper than 35\%, beyond which the inclusion of shallow mask lines has little effect (Fig.\ref{fig:maskcut}). This might indicate the intrinsic stability of the star (stellar jitter), since PARAS is stable to 1.5~m~s$^{-1}$ or better.

\section{Observations \& new RV Results on $\sigma$ Dra \& HD 9407}

Observational procedure is carefully established to ensure that all calibration exposures and target visibility information is available to enable uninterrupted execution of the data analysis pipeline. A nightly log is maintained to complement the FITS header information. The nightly calibration sequence includes 5 bias frames and 5 flat frames (both fibers illuminated with a tungsten lamp), and several ThAr-ThAr frames (both fibers illuminated with the calibration lamp) throughout the night. The purpose of the latter is to carefully measure absolute instrument drift, as well as differential drifts (Fig.\ref{fig:parasstability2}). Science observations are usually simultaneous star-ThAr exposures (2-3 exposures per night per target). Depending on the brightness of the target, these exposures are typically between 600 - 1200s. The science observations are typically bracketed by 300s ThAr-Thar calibration frames. In addition, we collect 600s and 1200s dark-ThAr exposures (i.e. with an illuminated calibration fiber but no light in the star fiber) every night before the start of the observations. These frames are used to mitigate the effects of bleeding of the ThAr lines that overlap with stellar spectra.

Every afternoon we also pump the spectrograph vacuum vessel to 0.07 mbar which rises to 0.13 mbar at dawn the following morning. Thus, the spectrograph always sees a pressure variation of 0.06 mbar during the night of observations. A pressure change of 0.06 mbar ideally would correspond to 6~m~s$^{-1}$ instrument absolute drift \citep{Pepe:2002}, however, this is overshadowed by larger absolute drifts of 90 - 150~m~s$^{-1}$ due to the temperature drifts of $\Delta$T $\sim$0.02$^{\circ}$C within the vacuum enclosure (Fig.\ref{fig:parasstability1}).

The quality and completeness of the order trace depends on the uniform illumination of the flat field exposure. We use a color balancing filter for uniform illumination from blue to red wavelengths since the flat lamp filament temperature is only 3000K. The color balancing filter, consisting of two Hoya LB200 filters, improves the flat field exposures greatly and enables us to extract all 99 existing orders for the A fiber. For the B fiber the first order on the blue side is only partially registered on the CCD and therefore we discard this order for both the fibers and effectively only extract 98 orders for science frames. This extends our spectral range to approximately 3800-9500\AA . However, the argon saturation in the red curtails the useful region at about 7000\AA . High precision polishing of the spectrograph optics makes sure that the scattering is minimal and, in principle, we can get precision RVs even in the presence of strong saturated Argon lines in the red region beyond 7000\AA. It is worth mentioning here that E2V advised us (private communication) that because of deep-depletion nature of the CCD it is not advisable to have a particular set of pixels on the CCD always super saturated (the position of strong Argon lines). Hence for RV work we use a long-wavelength cut-off filter at 7000\AA\ and currently only perform wavelength calibration upto 7000\AA. We do have plans of using ThNe or UNe \citep{Redman:2011} spectral lamps in the future to explore the wavelength region between 7000\AA\  and 9500\AA.

After several engineering runs between March 2011 and March 2012, we initiated our science observations from Spring-Summer 2012. The science program mainly consists of

\begin{itemize}
\item{radial velocity measurements around bright RV standard stars (stars brighter than 7th magnitude)}
\item{transit follow-up RV observation of some of the bright targets up to 9.5mag (in V-band)}
\end{itemize}

 The primary aim of observing bright RV standard stars is that of establishing long term RV precision of sub-2 ~m~s$^{-1}$ in the time period of several months to slightly longer than a year. This will ultimately enable us to look for short period (up to $\sim$ 1 month) super-Earths around bright targets up to 7th magnitude (photon noise limit of 1.25 ~m~s$^{-1}$ at 6.5 magnitude). For our transit follow-up program between 8th \& 10th magnitude stars, modest RV precision of $\sim$8~m~s$^{-1}$ (photon noise limit of about 4 ms$^{-1}$ at 8th magnitude) is obtained, which is sufficient for confirmation of hot Jupiter transiting candidates.

\begin{figure*}[] 
   \centering
   \includegraphics[width=0.75\textwidth]{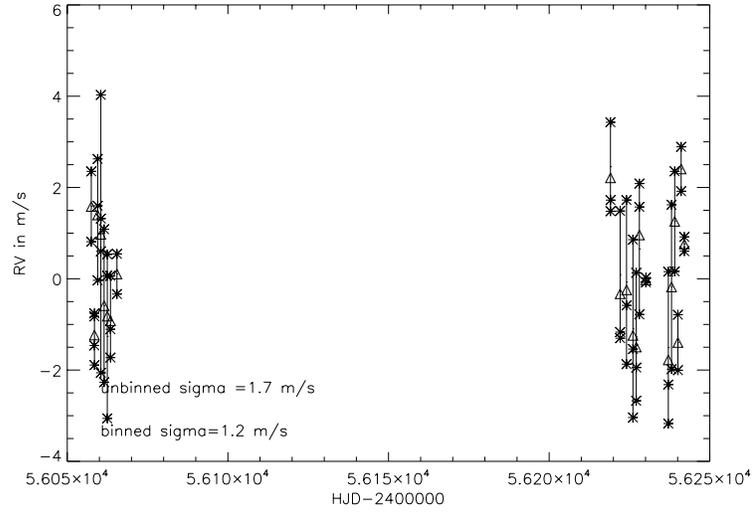}
    \caption{Mean subtracted radial velocity results for $\sigma$Dra over a period of 7 months from May 2012 to Nov 2012. The star is 4.8 mag in V-band, spectral type G9/K0V. The asterisks are the individual data points and the open-triangles are the velocities of the binned data (mean of the three visits per night) and the lines show the RV variations per night.}
   \label{fig:sigdra}
\end{figure*}

\begin{figure*}[] 
  \centering
  \includegraphics[width=0.75\textwidth]{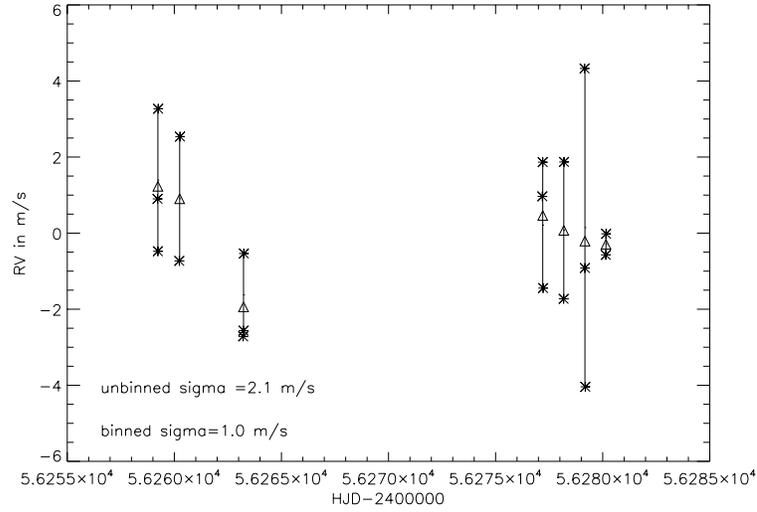}
   \caption{Mean subtracted radial velocity results for HD 9407 over a period of little over a month (Nov-Dec 2012). The star is 6.5mag in the V-band, spectral type G0V. The asterisks are the individual data points and the open-triangles are the velocities of the binned data (mean of the three visits per night) and the lines show the RV variations per night.}
  \label{fig:hd9407}
\end{figure*}

We observed the bright RV standard star, $\sigma$ Dra \citep{Wright:2008,Johnson:2006}, to demonstrate the capabilities of the instrument. Using a G2 stellar mask we achieve a velocity {\it rms} of 1.7~m~s$^{-1}$ for a period of $\sim$7 months (from May 2012 to November 2012; Fig. \ref{fig:sigdra}). We considered orders within a range of 3800-6900\AA . If the observations are binned nightly, the scatter reduces to 1.0~m~s$^{-1}$. The photon noise is computed from the full information content of the spectra, using the method prescribed by \citet{Bouchy:2001}. We estimate the fundamental photon noise limit using observations of $\sigma$ Dra, considering rotational broadening {\it v}sin{\it i} $\sim$ 1.5~km~s$^{-1}$, in the wavelength range of 3800-6900\AA~and with a spectrograph resolution of 67,000. This yields a value of 0.95~m~s$^{-1}$. We also observed HD 9407, a G0V Vmag $\sim$6.5, with expected photon noise error of 1.25~m~s$^{-1}$. Figure \ref{fig:hd9407} shows RV precisions of 2.1~m~s$^{-1}$ (nightly binned 1.1~m~s$^{-1}$) on the star HD 9407 for a period of $\sim$ 1 month.

\section{Future Upgrades to Enhance RV precision}
In the preceding sections, we have outlined the necessity of increasing the temperature control precision. While the inner room is well insulated with wood and PUF material (Polyurethane Foam), the vacuum vessel itself sits on an insulated and vibration-isolated concrete pier which takes the weight of the vessel. The points of contact between the vessel and the pier are through eight flat legs which rest on trisolators. The 1 foot $\times$1 foot $\times$ 4 inch thick trisolators act both as vibration isolators and thermal isolators. Our investigations suggest that there is some heat leakage from the insulated concrete pier on which the vacuum vessel sits. These leakages can cause vey serious temperature variations within the vessel if not taken care of. At present we have made arrangements by installing small fans and extra heaters around the eight legs of the vessel along with controlled heating pads. Typically $\Delta$T of $\sim$0.06$^{\circ}$C in the inner room yields about $\Delta$T of 0.02$^{\circ}$C inside the vessel (figure \ref{fig:parasstability1}), which in turn produces nightly absolute drifts of up to 160m/s. To reduce the absolute drifts to less than 50m/s we need to reduce the maximum variation in $\Delta$T from the current value of $\sim$0.06$^{\circ}$C to 0.02$^{\circ}$C in the inner room chamber and further isolate the vessel from heat leakage through the pier. We will perform further tests over a stretch of several months and develop an electronic system that will control the extra heaters in a differential manner, with the fan speed responding to sensors; thereby integrating the new elements with the existing temperature control system.

Another aspect that will help improve RV precision is optical scrambling via the introduction of octagonal fibers into fiber feed. Recent tests at the Geneva Observatory test bench on octagonal fibers \citep{Chazelas:2010} have shown significant stability of the near field, in response to changing illumination at the fiber input. Thus, a combination of octagonal and circular fibers are expected to provide a highly stable PSF at the spectrograph input. This helps mitigate errors due to poor telescope tracking, or seeing variations. \citet{Bouchy:2013} have demonstrated the PSF stability introduced by octagonal fibers with the SOPHIE spectrograph. After the installation of octagonal fibers, the authors have shown that they can achieve sub-2~m~s$^{-1}$ precision on stars where they were getting 5 to 10~m~s$^{-1}$ precision.

The PRL spectrograph (PARAS) has already demonstrated better than 2~m~s$^{-1}$ RV precision on select stars. In January 2013 we included a 2m long 50 $\mu$m core Ceramoptec octagonal fiber  coupled with the existing 20m long circular fiber. The star is focused on the octagonal fiber tip. This should generate a very stable PSF at the spectrograph and may produce sub-1~m~s$^{-1}$ precision. It may be also necessary to further improve the telescope tracking system and the temperature control of the spectrograph to realize consistent sub-1~m~s$^{-1}$ RV performance over time. Initial results are very encouraging and we plan to observe stable stars like $\tau$Ceti over a period of several months to demonstrate this.  These observations of $\tau$Ceti are scheduled to begin in October 2013. While we have focused largely on the instrument design and performance here, the results of the upgrades and improvements on radial velocity precision will be presented in the scientific publications resulting from the stabilized PARAS spectrograph.


\section*{Acknowledgements}
Eric Harvey Richardson, who was involved in the optical design of PARAS and its fiber feed passed away on November 20, 2011 at the age of 84 battling cancer. His contributions and influence have been felt throughout this work.

Funding for the PARAS project  is provided by the Department of Space, Government of India through the Physical Research Laboratory. We would like to express our gratitude to J.N. Goswami the director of PRL for making this project possible. We are grateful to Larry Ramsey (Penn State) \& Francesco Pepe (Geneva Observatory) for detailed discussions over the years on aspects of building precision RV spectrographs and also the Geneva Observatory for allowing us to use their stellar masks for RV reductions. The contribution of D. Subramanyam, \& his team from SAC, ISRO on initial Vacuum Chamber concept design and Finite-Element analysis verifications is highly appreciated. We also would like to thank Bhas Bapat (PRL) for discussion on the Vacuum Chamber design. We thank Venkat Ramani of Aditya High Vacuums for coming up with the innovative design for the optical Fiber feed-through, and for making many trips to Mt. Abu to solve the vacuum leak issues. We also thank N.M. Ashok (PRL) for generously granting us telescope time for on-sky engineering runs during the spectrograph testing phase and the Observatory staff at Mt. Abu for their support, and Rohit Deshpande (Penn State) for some early observations. We thank Robert Leach (SDSU) for his suggestions on reduction of the EM interference to reduce the CCD read noise at the Observatory. Finally we would like to thank the anonymous referee for his numerous good suggestions which improved the quality of the paper.

This work has been partially supported by funding from the Center for Exoplanets and Habitable Worlds. The Center for Exoplanets and Habitable Worlds is supported by the Pennsylvania State University, the Eberly College of Science, and the Pennsylvania Space Grant Consortium. SM also acknowledges support from NSF grant AST1006676, AST 1126413, PSARC, and the NASA Astrobiology Institute  pursuit of precision radial velocities in the NIR and optical.  This work made use of the SIMBAD database (operated at CDS, Strasbourg, France), NASA's Astrophysics Data System Bibliographic Services, and the NASA Star and Exoplanet Database (NStED).


\clearpage
\bibliographystyle{apj}

\clearpage

\begin{table*}[]
\fontsize{8}{10}\selectfont
\begin{center}
\caption{Surface data summary from Zemax for the spectrograph from the slit position onwards.}   
\label{table:ZE}
\begin{tabular}{ c  c  c  c  c  c  c }
 \hline
 \hline
Surf & Comment	& Radius	& Thickness	& Glass	& Diameter	& Conic   \\
\hline
1	& Slit Position	& Infinity	& 0	&	& 0.160	&     \\
2	& Coords. for M1	& -	& 1299.974	& -	& -	&    \\
3	& M1 (1)	& -2600	& -1222.825	& MIRROR	& 373.9709	& -1    \\
4	& Offset	&	& 	&	& 	&    \\
5	& Gamma	& 	& 0	& 	& 	&  \\
6	& +90deg Z-rotation	& 	& 0	& 	& 	&  \\
7	& Blaze angle (75.0deg)	&	& 0	&	&	&  \\
8	& R4 Echelle	& Infinity	& 0	& MIRROR	& 389.3651	& 0  \\
9	& -Blaze	&	& 0	&	&	&  \\
10	& -90deg Z-rotation	&	& 0	&	&	&  \\
11	& -Gamma 	&	& 0	&	&	&  \\
12	& -Offset	&	& 1222.825	&	&	&  \\
13	& M1 (2)	& -2600	& -1239.974	& MIRROR	& 371.0768	& -1 \\
14	& 	&	& 0  &	&   \\
15	& Fold Mirror	& Infinity	& 0	& MIRROR	& 156.0679	& 0  \\
16	&	& 	& 60	&	&	& \\
17	& Intermediate Focus	&	& 1301.647	&	&	& \\
18	& M2	& -2600	& 0	& MIRROR	& 450.0312	& -1 \\
19	& To cross-disperser; decenter in Y=142.01mm & & -1027.004 &  &	&  \\
20	& Tilt at prism; 59.43deg about X	&	& 0	&	&	&  \\
21	& Prism entrance	& Infinity	& 0	& PBM8Y	& 213.0499	& 0  \\
22	& Tilt in X by -32.80deg	& 	& -122	&	&	&   \\
23	& Tilt in X by -32.80deg  &	& 0	&	&	&   \\
24	& Prism exit	& Infinity	& 0	& 	& 208.0738	& 0  \\
25	& Tilt in X by -59.43deg	&	& 0	&	&	&  \\
26	& Orientate1	&	&  -125	& 	& 	&  \\
27	& Orientate2	&	& 0	&	&	&  \\
28	& Start of camera lens system	& -895.18	& -15.5	& S-FPL51	& 133.3989	&  0  \\
29	&	& Infinity	& -29.24854	&	& 133.3989	& 0   \\
30	&	& -744.61	& -15	& PBM8Y	& 156	&  0   \\
31	& Soft cement	& -1141.8	& -0.03	& RTV141	& 156	&  0  \\
32	&	& -1141.8	& -21.5	& S-FPL51	& 156	&  0  \\
33	& Soft cement	& 480.16	& -0.03	& RTV141	& 156	&  0   \\
34	&	& 480.16	& -11.5	& PBM8Y	& 156	&  0   \\
35	&	& Infinity	& -507.9644	&	& 156	&  0   \\
36	&	& -263.33	& -12.8	& S-FPL51	& 150	&  0   \\
37	& Soft cement	& -410.83	& -0.03	& RTV141	& 150	&  0  \\
38	&	& -410.83	& -16	& PBM8Y	& 150	&  0  \\
39	& Camera exit	& Infinity	& -148.6354	&	& 150	&  0  \\
40	& Orientate3	&	& 0	&	&	&   \\
41	& Top of Shutter	& Infinity	& -27.94	&	& 81.54351	&  0   \\
42	& Window	& Infinity	& -6.35	& FuseSILICA	& 78.73493	& 0   \\
43	& Window exit	& Infinity	& -9	&	& 78.30927	& 0  \\
44	& Detector	& Infinity	& 	&	& 77.44268	& 0  \\
\hline
\end{tabular}
\end{center}
$\bf{Notes}$: M1 and M2 are the off-axis parabolic mirrors. The Offset in surface 6 is decenter in Y = -136.27mm, in surface 5, the Gamma = 0.45deg (tilt about X); in surface 20, the tilt at Prism is 59.43deg about X; in surface 26, Orientate1 is 5.07deg about X; in surface 27, Orientate2 are decenter in Y=-11.42mm, tilt about X = -3.34deg, tilt about Y = 0.39deg; in surface 40 Orientate3 are decenter in X = 6.86mm, decenter in Y = -3.82mm, tilt in X = 5.67deg, tilt in Y = -0.41deg. Soft cement is optical glue called RTV 141 A \& B. It is a two component, poly-addition reaction, room temperature curing silicone elastomer manufactured by BlueStar Silicones.
\end{table*}
\clearpage
\begin{table}[]
\begin{center}
\caption{Measured Quantum Efficiency of the PARAS Science CCD Array; measurement done by E2V}
\label{table:QE}
\begin{tabular}{ c  c }
 \hline
 \hline
  Wavelength (\AA) & Quantum Efficiency (QE)  \\
  \hline
  3500 & 57.2\%  \\
  4000 & 92.4\%  \\
  5000 & 96.5\%  \\
  6500 & 87.9\%  \\
  9000 & 51.9\%  \\
  \hline
\end{tabular}
\end{center}
\end{table}

\newpage


\end{document}